\documentclass[12pt]{article}
\usepackage{hyperref}
\usepackage{amsmath}
\usepackage{amssymb,amsthm}
\usepackage[english]{babel}
\usepackage[textwidth=18cm,textheight=22cm]{geometry}
\newcommand{\C}{\mathbb{C}}
\newcommand{\R}{\mathbb{R}}
\newcommand{\Z}{\mathbb{Z}}
\newcommand{\g}{\mathfrak{g}}

\renewcommand{\d}{\mathrm{d}}

\newcommand{\ad}{\operatorname{ad}} \newcommand{\Ad}{\operatorname{Ad}}
\newtheorem{pro}{Proposition}

\begin{document}
%\section{The factorization problem for canonical transformations
%and  dispersionless hierarchies}
\title{On the $r$-th dispersionless Toda hierarchy I:\\ Factorization problem, symmetries
 and some solutions}

\author{Manuel Ma{\~n}as\\ Departamento de F{\'\i}sica Te{\'o}rica II, Universidad Complutense\\ 28040-Madrid, Spain\\
email: manuel@darboux.fis.ucm.es}

\maketitle

\abstract{For a family of Poisson algebras, parametrized by
$r\in\Z$, and  an associated Lie algebraic splitting,  we consider
the factorization of given canonical transformations. In this
context we rederive the recently found $r$-th dispersionless
modified KP hierachies and $r$-th dispersionless Dym hierarchies,
giving a new Miura map among them. We also found a new integrable
hierarchy which we call the $r$-th dispersionless Toda hierarchy.
Moreover, additional symmetries for these hierarchies are studied
in detail and new symmetries depending on arbitrary functions are
explicitly constructed for the $r$-th dispersionless KP, $r$-th
dispersionless Dym and $r$-th dispersionless Toda equations.  Some
solutions are derived by examining the imposition of a time
invariance to the potential $r$-th dispersionless Dym equation,
for which a complete integral is presented and therefore an
appropriate envelope leads to a general solution. Symmetries and
Miura maps are applied to get new solutions and solutions of the
$r$-th dispesionless modified KP equation.}

\section{Introduction}

The study of dispersionless integrable hierarchies is a subject of
increasing activity in the theory of integrable systems. This was
originated in several sources, let us metion mention here the
pioneering work of Kodama and Gibbons \cite{kodama} on the
dispersionless KP, of Kupershimdt on the dispersionless modified
KP \cite{ku} and the role of Riemann invariants and hodograph
transformations found by Tsarev \cite{tsarev}. The important work
of Takasaki and Takabe, \cite{takasaki-takebe},
\cite{takasaki-takebe_toda} and \cite{takasaki-takebe-3} which
gave the Lax formalism, additional symmetries, twistor formulation
of the dispersionless KP and dispersionless Toda hierarchies. For
the dispersionless Dym (or Harry Dym) equation see \cite{Li}. The
appearance of dispersionless systems in topological field theories
is also an important issue, see \cite{krichever} and
\cite{dubrovin}. More recent progress appears in relation with the
theory of conformal maps \cite{gibbons} and \cite{wiegman},
quasiconformal maps and $\bar\partial$- formulation
\cite{konopelchenko}, reductions of several type \cite{otros} and
\cite{guil-manas-martinez}, additional symmetries
\cite{martinez-manas} and twistor equations \cite{Previo}, on
hodograph equations for the Boyer--Finley equation
\cite{works_on_boyer} and its applications in General Relativity,
see also \cite{dunajski} and \cite{dunajski2}.  It is also
remarkable the approach given in \cite{ferapontov} to the theory.
Finally, we comment the contribution on electrodynamics and the
dispersionless Veselov--Novikov equation
\cite{konopelchenko-moro}.

Recently, a new Poisson bracket and associated Lie algebra
splitting, therefore using a Lax formalism and an $r$-matrix
approach, was presented in \cite{blaszak} to construct new
dispersionless integrable hierarchies and latter on, see
\cite{blaszak2}, the theory was further extended. We must remark
that since the work \cite{go} and \cite{Li} it was known the
possibility for a $r$-matrix formulation of dispersionless
integrable systems.

In this paper we shall use this new splitting together with a
standard technique in the theory of integrable
 hierarchies, the factorization problem, to get a new dispersionless
integrable hierarchy, which we call the $r$-th dispersionless Toda
hierarchy. For that aim we consider the Lie group of canonical
transformations associated with a particular Poisson bracket
together with a factorization problem induced by a $r$-matrix
associated with a canonical splitting of the corresponding algebra
of symplectic vector fields \cite{blaszak}. These new hierarchies
contains dispersionless integrable equations derived previously in
\cite{blaszak} ---which we call $r$-th dispersionless modified KP
and $r$-th dispersionless Dym equations (for $r=0$ we get the
well-known dispersionless modified KP and dispersionless Dym
equations)---. However, we found new integrable hierarchies, which
we decided to name as $r$-th dispersionless Toda hierarchy (as for
$r=1$ we get the well-known dispersionless Toda hierarchy).

In \cite{chen-tu} the Miura map among the dispersionless modified
KP and dispersionless Dym equations was presented, here we extend
those results to the present context.  We also study in detail the
additional symmetries of these hierarchies and in particular for
the three integrable equations: $r$-th dispersionless modified KP,
$r$-th dispersionless Dym and $r$-th dispersionless Toda equations
we get new explicit symmetries depending on arbitrary functions,
in the spirit of the symmetries given in \cite{dunajski} for the
dispersionless KP equation.

Latter  we find a complete integral for the $t_2$-reduction of the
potential $r$-th dispersionless Dym equation. Thus, using the
method of the complete solution, an appropriate envelope leads to
its general solution. Then, when the symmetries are applied we get
more general solutions, non-$t_2$ invariant of the potential
$r$-th dispersionless Dym equation. Using the Miura map new
solutions of the $r$-th dispersionless modified KP equation are
gotten and the corresponding functional symmetries are applied to
get more general families of solutions. Finally, we also derive
from the factorization problem twistor equations for these
integrable hierarchies.

The layout of this paper is as follows. In \S 2 we consider the
Lie algebra setting, the factorization problem and its
differential description, Lax functions and Zakharov--Shabat
formulation are given, as well. Then, in \S 3 the corresponding
integrable hierarchies are derived together with the Miura map.
Next, in \S 4, we introduce Orlov functions and show how the
factorization problem constitutes a simple framework to derive the
corresponding additional symmetries of the integrable hierarchies.
In particular, additional symmetries (which depend on arbritrary
functions of $t_2$) of the integrable dispersionless equations
discussed in \S 3 are found. Finally, \S 5 is devoted to some
solutions of these integrable hierarchies and \S 6 to a twistor
formulation of these dispersionless integrable hierarchies derived
from the factorization problem. In a forthcoming paper we will
give a twistor formulation of these hierarchies which is derived
independently of the factorization problem.

\section{Factorization problem and its differential versions}
We shall work with the Lie algebra $\g$ of Laurent series
$H(p,x):=\sum_{n\in\Z}u_n(x)p^n$ in the variable $p\in\R$ with
coefficients depending on the variable $x\in\R$, with Lie
commutator given by the following Poisson bracket \cite{blaszak}
\[
\{H_1,H_2\}=p^r\Big(\frac{\partial H_1}{\partial p}\frac{\partial
H_2}{\partial x}-\frac{\partial H_1}{\partial x}\frac{\partial
H_2}{\partial p}\Big),\quad r\in\Z.
\]

Observe that for each $r\in \Z$ we are dealing with a different
Lie algebra; notice also that this Poisson bracket is associated
with the the following sympletic form
\[
\omega:=p^{-r}\d p\wedge\d x.
\]

\subsection{The Lie algebra splitting}
We shall use the following triangular type splitting of $\g$ into
Lie subalgebras
\begin{equation}\label{splitting}
\g=\g_>\oplus\g_{1-r}\oplus \g_<
\end{equation}
where
\begin{equation*}
\g_\gtrless:=\C\{u_n(x)p^n\}_{n\gtrless(1-r)},\quad
\g_{1-r}:=\C\{u(x)p^{1-r}\},
\end{equation*}
and therefore fulfil  the following property
\[
\{\g_\gtrless,\g_{1-r}\}= \g_\gtrless.
\]
If we define the Lie subalgebra $\g_\geqslant$ as
\[
\g_\geqslant:=\g_{1-r}\oplus\g_>
\]
we have  the  direct sum decomposition of the Lie algebra $\g$
given by
\begin{equation}\label{algebra-factor}
\g=\g_<\oplus\g_\geqslant.
 \end{equation}

We remark that allowing the variable $p$ to take values in $\C$ we
have the following interpretation for the above splitting
\eqref{algebra-factor}. Suppose that $\g$ is the set of analytic
functions in some annulus of $p=0$. Then, $H\in\g_<$ iff
$H(p)p^{-1+r}$ is an analytic function outside the annulus which
vanish at $p=\infty$. On the opposite a function
$H\in\g_\geqslant$ if $H(p)p^{-1+r}$  has an analytic extension
inside the annulus.

%\psfrag{a}{$\g_<$: analytic extension of $p^{1-r}H$ normalized to
%1 at $\infty$}\psfrag{b}{$\g_\geqslant$: analytic extension of
%$p^{1-r}H$}\psfrag{g}{$\g$}\psfrag{0}{$p=0$}
%\begin{center}
%\includegraphics[width=8cm]{annulus.ps}
%\end{center}

Observe that the induced Lie commutator in $\g_{1-r}$ is
\[
\{f(x)p^{1-r},g(x)p^{1-r}\}=(1-r)W(f,g),
\]
where $W(f,g):=fg_x-gf_x$ is the Wro\'{n}skian of $f$ and $g$.
Only when $r=1$ the Lie subalgebra $\g_{1-r}$ is an Abelian Lie
subalgebra.

An alternative realization of $\g$ is through the adjoint action
\[
\begin{aligned}
 \ad&: & \g & &\rightarrow& \mathfrak{X}(\R^2),\\
    & & H & & \mapsto &  \ad_H:=\{H,\cdot\}.
     \end{aligned}
\]
so that
\[
\ad_H=p^r\frac{\partial H}{\partial p}\frac{\partial}{\partial x}-
p^r\frac{\partial H}{\partial x}\frac{\partial}{\partial p}.
\]

\subsection{The factorization problem} Given an element $H\in\g$
the corresponding vector field $\ad_H$ generates a symplectic
diffeormorphism (canonical transformation) $\Phi_H$,  given by
\[
\Phi_H:=\sum_{n=0}^\infty\frac{1}{n!}(\ad_H)^n.
\]
This transformation corresponds to the element $h=\exp(H)$
belonging to the  local Lie group $G$ generated by $\g$. In fact,
$\Phi_H=\Ad_h$, the adjoint action of the Lie group $G$ on $\g$.

In what follows we shall denote by $G_<,G_{1-r},G_>$ and
$G_\geqslant$ the local Lie groups corresponding to the Lie
algebras $\g_<,\g_{1-r},\g_>$ and $\g_\geqslant$, respectively.

 Given $h,\bar h\in G$ in the local Lie group $G$
the finding of $h_<\in G_<$ and $h_\geqslant\in G_\geqslant$ such
that the following factorization holds
\begin{equation}\label{pre-factorization}
h_<\cdot h=h_\geqslant\cdot\bar h,
 \end{equation}
 will play a pivotal role in what follows.

Furthermore, given  two sets of deformation parameters
$(t_1,t_2,\dots)$ and $(\bar t_1,\bar t_2,\dots))$ and
 corresponding elements in the Lie algebra $\g$
\[
t(p):=t_1p^{2-r}+t_2 p^{3-r}+\cdots\in\g_>,\quad \bar t(p):=\bar
t_1p^{-r}+\bar t_2 p^{-r-1}+\cdots\in\g_<,
\]
we shall analyze the following deformation of
\eqref{pre-factorization}:
 \begin{equation}\label{factorization}
\psi_<\cdot \exp(t)\cdot h=\psi_\geqslant\cdot\exp(\bar
t)\cdot\bar h.
 \end{equation}

 Notice that there is no loss of generality if we set $\bar h=1$
 in \eqref{factorization}  so that
 \begin{equation}\label{factorization-simple}
\exp(t)\cdot h\cdot\exp(-\bar t)=\psi_<^{-1}\cdot \psi_\geqslant.
 \end{equation}

\subsection{Differential consequences of the factorization problem}

 A possible way to study \eqref{factorization} is by analyzing its
 differential versions; i. e., by taking right derivatives. Given
 a derivation $\partial$ of a Lie algebra $\g$ one defines the
 corresponding right-derivative in the associated local Lie group
 by
 \begin{equation}\label{right-derivative}
 \partial h\cdot h^{-1}:=\sum_{n=0}^\infty\frac{1}{(n+1)!}(\ad
 H)^n(\partial H),\quad h:=\exp(H),\quad H\in\g.
 \end{equation}

Hence, by taking right-derivatives of \eqref{factorization} with
respect to
\[
\partial_n:=\frac{\partial}{\partial t_n},\quad
\bar\partial_n:=\frac{\partial}{\partial \bar t_n},\quad
\]
we get
\begin{align}
\partial_n\psi_<\cdot\psi_<^{-1}+\Ad_{\psi_<}(p^{n+1-r})&=\partial_n\psi_\geqslant\cdot
\psi_\geqslant^{-1},\label{factorization-diff-1}\\
\bar\partial_n\psi_<\cdot\psi_<^{-1}&=\bar\partial_n\psi_\geqslant\cdot
\psi_\geqslant^{-1}+\Ad_{\psi_\geqslant}(p^{1-r-n}).\label{factorization-diff-2}
\end{align}
If we further factorize
\[
\psi_\geqslant=\psi_{1-r}\cdot\psi_>
\]
equation \eqref{factorization-diff-1} decomposes --according with
\eqref{splitting}-- in the following three equations
\begin{align}
\partial_n\psi_<\cdot\psi_<^{-1}+P_<\Ad_{\psi_<}p^{n+1-r}&=0,\label{<}\\
P_{1-r}\Ad_{\psi_<}p^{n+1-r}&=\partial_n\psi_{1-r}\cdot\psi_{1-r}^{-1},\label{1-r}\\
P_>\Ad_{\psi_<}p^{n+1-r}&=\Ad_{\psi_{1-r}}\big(\partial_n\psi_>\cdot\psi_>^{-1}\big).\label{>}
\end{align}
Similar considerations applied to \eqref{factorization-diff-2}
lead to
\begin{align}
\bar\partial_n\psi_<\cdot\psi_<^{-1}&=\Ad_{\psi_{1-r}}P_<\Ad_{\psi_>}p^{1-r-n},\label{bar<}\\
0&=\bar\partial_n\psi_{1-r}\cdot\psi_{1-r}^{-1}+
\Ad_{\psi_{1-r}}P_{1-r}\Ad_{\psi_>}p^{1-r-n},\label{bar1-r}
\\
0&=\bar\partial_n\psi_>\cdot\psi_>^{-1}+P_>\Ad_{\psi_>}p^{1-r-n}.\label{bar>}
\end{align}

Now we shall show that we can interchange the roles of $\g_<$ and
$\g_\geqslant$. For this aim we introduce the map
\begin{equation}\label{intertwin}
\begin{aligned}F&:\R^2&\rightarrow&\R^2&\\
&(p,x)&\mapsto& (p':=1/p,x'=-x)&
\end{aligned}
\end{equation}
that together with $r'=2-r$ may be considered as a ``canonical"
transformation in the sense that the canonical form
$\Omega=p^{-r}\d p\wedge\d x$ transforms onto
\[
\Omega'=(p')^{-r'}\d p'\wedge\d x'.
\]
If we  denote
\[
\g'_\gtrless:=\C\{u_n(x)(p')^n\}_{n\gtrless(1-r')},\quad
\g'_{1-r'}:=\C\{u(x)p^{1-r'}\},
\]
then
\[
\g_>\to \g'_<,\quad \g_{1-r}\to \g'_{1-r'},\quad \g_<\to\g'_>.
\]
%This implies for our Lie algebra splitting the following
%transformation
%\[
%\g_{<}\oplus \g_{\geqslant}\to \g'_{\leqslant}\oplus\g'_{>}
%\]

Observe that
\[
\begin{aligned}
t(p)&=t_1p^{2-r}+t_2 p^{3-r}+\cdots\to t_1\,(p')^{-r'}+ t_2\,
(p')^{-r'-1}+\cdots=\bar t'(p'),
\\
\bar t(p)&=\bar t_1p^{-r}+\bar t_2 p^{-r-1}+\cdots\to \bar
t_1\,(p')^{2-r'}+\bar t_2 \,(p')^{3-r'}+\cdots=t'(p') ,
\end{aligned}
\]
Thus, $t_n'=\bar t_n$ and $\bar t_n'=t_n$. Finally, the
factorization \eqref{factorization} transforms as
\begin{equation}%\label{factorization}
\psi_<\cdot\exp(t)\cdot h=\psi_{1-r}\cdot\psi_>\cdot\exp(\bar
t)\cdot\bar h \to \psi'^{-1}_{1-r'}\cdot\psi'_>\cdot\exp(\bar
t')\cdot h=\psi'_<\cdot\exp(\bar t)\cdot\bar h
 \end{equation}
 With these observations
is easy to see that our map transforms equations \eqref{<},
\eqref{1-r} and \eqref{>} into \eqref{bar>}, \eqref{bar1-r} and
\eqref{bar<}, respectively.

\subsection{Lax formalism and Zakharov--Shabat representation}
Equations \eqref{<}-\eqref{bar>} can be given a Lax form, for that
aim we first introduce the following Lax functions
\begin{equation}\label{lax_def}
\begin{aligned}
L&:=\Ad_{\psi_{<}}p,\\
\bar\ell&:=\Ad_{\psi_>}p,\\
\bar L&:=\Ad_{\psi_{1-r}}\bar\ell=\Ad_{\psi_{\geqslant}}p
\end{aligned}
\end{equation}
in terms of which equations \eqref{factorization-diff-1} and
\eqref{factorization-diff-2} read as
\begin{align*}
\partial_n\psi_<\cdot\psi_<^{-1}&=\partial_n\psi_\geqslant\cdot
\psi_\geqslant^{-1}-L^{n+1-r},\\
\bar\partial_n\psi_\geqslant\cdot
\psi_\geqslant^{-1}&=\bar\partial_n\psi_<\cdot\psi_<^{-1}-\bar
L^{1-r-n}.
\end{align*}

So that
\begin{equation}\label{L}
\begin{aligned}
\partial_n\psi_<\cdot\psi_<^{-1}&=-P_<L^{n+1-r},&
\partial_n\psi_{\geqslant}\cdot\psi_{\geqslant}^{-1}&=P_{\geqslant}L^{n+1-r},\\
\bar\partial_n\psi_<\cdot\psi_<^{-1}&=P_<\bar L^{1-r-n},&
\bar\partial_n\psi_{\geqslant}\cdot\psi_{\geqslant}^{-1}&=-P_{\geqslant}\bar
L^{1-r-n},
\end{aligned}
\end{equation}
and therefore the following Lax equations hold
\begin{equation}\label{lax}
\begin{aligned}
\partial_n L&=\{-P_<L^{n+1-r},L\},&
\partial_n\bar L&=\{P_{\geqslant}L^{n+1-r},\bar L\},\\
\bar\partial_n L&=\{P_<\bar L^{1-r-n},L\},&
 \bar\partial_n\bar L&=\{-P_{\geqslant}\bar L^{1-r-n},\bar L\}.
\end{aligned}
\end{equation}
To deduce these equations  just recall that if $B=\Ad_\phi b$, and
$\partial$ is a Lie algebra derivation, then $\partial
B=\{\partial\phi\cdot\phi^{-1},B\}+\Ad_\phi\partial b$.

For a further analysis \eqref{L} is essential to get the powers of
$L$ and $\bar L$. In the following proposition we shall show how
the powers of the Lax functions are connected with $\Psi_<$,
$\Psi_>$ and $\xi$, the infinitesimal generators of $\psi_<$,
$\psi_>$ and $\psi_{1-r}$, respectively
\begin{equation}\label{psis}
\begin{aligned}
\psi_<&:=\exp(\Psi_<),\quad \Psi_<:=\Psi_1 p^{-r}+\Psi_2
p^{-r-1}+\cdots,\\
\psi_>&:=\exp(\Psi_>),\quad \Psi_>:=\bar\Psi_1 p^{2-r}+\bar\Psi_2
p^{3-r}+\cdots,\\
\psi_{1-r}&:=\exp(\xi p^{1-r}).
\end{aligned}
\end{equation}

\begin{pro}\label{powers}
We can parameterize $L^m$, $\bar\ell^m$ and $\bar L^m$ in terms of
$\Psi_n$, $\bar\Psi_n$ and $\xi$ as follows
\begin{align}
\label{Lm}
L^m=&p^m+u_{m,0}p^{m-1}+u_{m,1}p^{m-2}+u_{m,2}p^{m-3}+O(p^{m-4}),\quad p\to\infty\\
\bar\ell^m=&p^m+\bar
 v_{m,0}p^{m+1}+\bar v_{m,1}p^{m+2}+\bar
 v_{m,2}p^{m+3}+O(p^{m+4}),\quad p\to 0,\\
\bar L^m=&\bar u_{m,-1} p^m+\bar u_{m,0}p^{m+1}+\bar
u_{m,1}p^{m+2}+\bar u_{m,2}p^{m+3}+O(p^{m+4}),\quad p\to 0
\end{align}
where  the first coefficients are
\begin{equation}
\label{u}\begin{aligned}
u_{m,0}=&-m\Psi_{1,x},\\
u_{m,1}=&m\Big(-\Psi_{2,x}+\frac{1}{2}\big(r\Psi_1\Psi_{1,xx}+(m-1)\Psi_{1,x}^2\big)
\Big),\\
u_{m,2}=&m\Big(-\Psi_{3,x}+\frac{1}{2}\big((r+1)\Psi_2\Psi_{1,xx}+
(2m-3)\Psi_{1,x}\Psi_{2,x}+r\Psi_1\Psi_{2,xx}\big)\\&-
\frac{1}{6}(r^2\Psi_1^2\Psi_{1,xxx}
+r(r+3m-4)\Psi_1\Psi_{1,x}\Psi_{1,xx} +(m-1)(m-2)\Psi_{1,x}^3)
\Big),
\end{aligned}
\end{equation}
\begin{equation}
\label{barv}
\begin{aligned}
 \bar v_{m,0}=&-m\bar\Psi_{1,x},\\ \bar
v_{m,1}=&m\Big(-\bar\Psi_{2,x}-\frac{1}{2}((2-r)
\bar\Psi_1\bar\Psi_{1,xx}
-(m+1)\bar\Psi_x^2\Big),\\
\bar
v_{m,2}=&m\Big(-\bar\Psi_{3,x}-\frac{1}{2}\big((3-r)\bar\Psi_2\bar\Psi_{1,xx}
-(2m+3)\Psi_{1,x}\Psi_{2,x}+(2-r)\bar\Psi_1\bar\Psi_{2,xx}\big)\\&-
\frac{1}{6}((2-r)^2\bar\Psi_1^2\bar\Psi_{1,xxx}
+(2-r)(r+3m+2)\bar\Psi_1\bar\Psi_{1,x}\bar\Psi_{1,xx}\\
&\hspace{7cm}+(m+1)(m+2)\bar\Psi_{1,x}^3) \Big),
\end{aligned}\end{equation}
and
\begin{equation*}\bar
u_{m,j}=\left\{\begin{aligned}&\Big(\dfrac{\xi\circ
X}{\xi}\Big)^{-\frac{m+1+j}{1-r}}\bar v_{m,j}\circ X,&
\int_x^X\frac{\d x}{\xi(x)}&=1-r, & &r\neq 1,\\
&\exp(-(m+1+j)\xi_x)\bar v_{m,j},&& && r=1.
\end{aligned}\right.
\end{equation*}
\end{pro}

\begin{proof}
To evaluate $L^m$ we recall that $L^m$ is connected to $p^m$
through a canonical transformation by
\[
L^m:=\Ad\psi_<p^{m}=p^{m}+\{\Psi_<,p^{m}\}+\frac{1}{2}\{\Psi_<,\{\Psi_<,p^{m}\}\}+
\frac{1}{6}\{\Psi_<,\{\Psi_<,\{\Psi_<,p^{m}\}\}\}+\cdots,
\]
and compute the first Poisson brackets to obtain
\begin{align*}
&\{\Psi_<,p^{m}\}=-m\Psi_{1,x}p^{m-1}-m\Psi_{2,x}p^{m-2}-m\Psi_{3,x}p^{m-3}+{O}(p^{m-4}),\\[10pt]
&\{\Psi_<,\{\Psi_<,p^{m}\}\}=m\big(r\Psi_1\Psi_{1,xx}+(m-1)\Psi_{1,x}^2\big)p^{m-2}
\\&\quad\quad\quad+
m\big((r+1)\Psi_2\Psi_{1,xx}+(2m-3)\Psi_{1,x}\Psi_{2,x}+r\Psi_1\Psi_{2,xx}\big)p^{m-3}
+{O}(p^{m-4}),\\[10pt]
&\{\Psi_<,\{\Psi_<,\{\Psi_<,p^{m}\}\}\}=-m(r^2\Psi_1^2\Psi_{1,xxx}
+r(r+3m-4)\Psi_1\Psi_{1,x}\Psi_{1,xx}\\&%%
\quad\quad\quad+(m-1)(m-2)\Psi_{1,x}^3)p^{m-3}+{O}(p^{m-4})
\end{align*}
when $p\to\infty$. Summing all terms and collecting those with the
same power on $p$ we get \eqref{u}. For $\bar\ell$ we use
\begin{multline*}
\bar
\ell^m:=\Ad\psi_>p^{m}=p^{m}+\{\Psi_>,p^{m}\}+\frac{1}{2}\{\Psi_>,\{\Psi_>,p^{m}\}\}+
\frac{1}{6}\{\Psi_>,\{\Psi_>,\{\Psi_>,p^{m}\}\}\}+\cdots,
\end{multline*}
and we get formulae \eqref{barv}. An alternative way to deduce
this is to use the intertwining transformation \eqref{intertwin},
$(p,x)\to (1/p,-x)$ together $r\to 2-r$, that intertwines $\g_>$
and $\g_<$. Thus, the expressions for $\bar v_{m,j}$ are obtained
from those for $u_{-m,j}$ by replacing $\partial_x^j\Psi_k$ by
$(-1)^j\partial_x^j\bar\Psi_k$ and $r$ by $2-r$.

As $\bar L^m=\Ad_{\psi_{1-r}}\bar\ell^m$ and we have already
obtained the expansion  of $\bar \ell$ in powers of $p$, in order
to compute $\bar L^m$, we need to characterize $\Ad_{\exp(\xi
p^{1-r})}(\phi(x)p^n)$; i. e., to characterize the canonical
transformation generated by $\xi p^{1-r}$. For that aim is useful
to perform the following calculation
\[
\ad_{\xi p^{1-r}}(\phi(x)p^n)=\{\xi
p^{1-r},\phi(x)p^n\}=\big([(1-r)\xi\partial_x-n\xi_x](\phi)\big)p^n
\]
so that
\[ \Ad_{\exp(\xi
p^{1-r})}(\phi(x)p^n)=[\exp((1-r)\xi\partial_x-n\xi_x)\phi(x)]p^n.
\]

For $r=1$; i.e., when $\g_{1-r}$ is an Abelian Lie subalgebra, the
action is easy to compute
\[ \Ad_{\exp\xi}(\phi(x)p^n)=\exp(-n\xi_x)\phi(x)p^n.
\]
However, for $r\neq 1$ the situation is rather more involved. Let
us analyze this non-Abelian situation. The function
\[
\Phi:=\exp(\lambda((1-r)\xi\partial_x-n\xi_x)\phi(x)
\]
is characterize --in a unique manner-- by the following initial
condition problem for a first-order linear PDE
\[
\begin{gathered}
\partial_\lambda \Phi=(1-r)\xi\partial_x \Phi-n\xi_x \Phi,\\
\Phi\big|_{\lambda=0}=\phi.
\end{gathered}
\]
The general solution of the PDE is
\[
\Phi=g(\lambda+c(x))\xi^{\frac{n}{1-r}},\quad
c(x):=\frac{1}{1-r}\int^x\frac{\d x}{\xi(x)}
\]
where $g$ is an arbitrary function, which is determined by the
initial condition
\[
\phi(x)=g(c(x))\xi^{\frac{n}{1-r}}.
\]
A better characterization of $g$ appears as follows: we first look
for a function $X(x)$  defined implicitly by the relation
\[
\int_x^X\frac{\d x}{\xi (x)}=(1-r)\lambda,
\]
and hence
\[
c(X)=\lambda+c(x).
\]
Now, taking into account the initial condition we deduce
\[
g(c(x))=\phi(x)\xi(x)^{\frac{n}{1-r}}
\]
so that
\[
g(c(X))=\phi(X)\xi(X)^{-\frac{n}{1-r}}
\]
and, consequently,
\[
\Phi(\lambda,x)=\phi(X(x))\Big(\frac{\xi(X(x))}{\xi(x)}\Big)^{-\frac{n}{1-r}}.
\]
 Hence
\begin{equation}\label{Xpsi}
\Ad_{\exp(\xi p^{1-r})}(\phi(x)p^n)=\left\{\begin{aligned}
&\Big(\dfrac{\xi(X(x))}{\xi(x)}\Big)^{-\frac{n}{1-r}}\phi(X(x))
p^n,&\int_x^X\frac{\d
x}{\xi (x)}&=(1-r), & &r\neq 1,\\
&\exp(-n\xi_x(x))\phi(x)p^n,&& & &r=1.
\end{aligned}\right.
\end{equation}
\end{proof}

 Observe that from the proof of the above Proposition \ref{powers} we deduce
that $u_{m,j}=-m\Psi_{j+1,x}+U_{m,j}$ where $U_{m,j}$ is a
nonlinear function of $\Psi_1$,\dots,$\Psi_{j}$ and its
$x$-derivatives.

To give the Lax equations \eqref{lax} the Zakharov--Shabat form;
i. e., a zero-curvature representation we introduce the the
exterior differential with respect to the variables $\{t_n,\bar
t_n\}_{n\geq 1}$
\[
\d:=\sum_{n\geq 1} \Big(\frac{\partial}{\partial t_n}\d
t_n+\frac{\partial}{\partial \bar t_n}\d \bar t_n\Big).
\]
Then, the factorization problem \eqref{factorization} implies
\begin{equation}\label{zero_curvature_1}
\d\psi_<\cdot\psi_<^{-1}+\Ad_{\psi_<}\d
t=\d\psi_{\geqslant}\cdot\psi_{\geqslant}^{-1}+\Ad_{\psi_\geqslant}\d
\bar t.
\end{equation}
and, if we define
\[
\Omega:=\Ad_{\psi_<}\d t-\Ad_{\psi_\geqslant}\d \bar t=\sum_{n\geq
1} (L^{n+1-r}\d t_n-\bar L^{1-r-n}\d \bar t_n),
\]
we may rewrite \eqref{zero_curvature_1} as follows
\begin{equation}\label{zero_curvature_2}
\Omega=\d\psi_{\geqslant}\cdot\psi_{\geqslant}^{-1}-\d\psi_<\cdot\psi_<^{-1}.
\end{equation}
 Equation \eqref{zero_curvature_2}  implies
\[
\d\psi_<\cdot\psi_<^{-1}=-P_<\Omega=:-\Omega_<,\quad
\d\psi_{\geqslant}\cdot\psi_{\geqslant}^{-1}=P_{\geqslant}\Omega=:\Omega_{\geqslant},
\]
that coincides with \eqref{L} when splitted in coordinates. Hence,
we deduce the following zero-curvature conditions
\[
\d\Omega_<=-\{\Omega_<,\Omega_<\},\quad
\d\Omega_{\geqslant}=\{\Omega_{\geqslant},\Omega_{\geqslant}\}.
\]

\section{The associated dispersionless integrable hierarchies}

We now deduce the integrable hierarchies associated with the
factorization problem \eqref{factorization}, namely the $r$-th
dispersionless modified KP, $r$-th dispersionless Dym and $r$-th
dispersionless Toda hierarchies.

\subsection{The $r$-th dispersionless modified KP hierarchy}
%On the Blaszak family of  ``modified dispersionless  KP" type hierarchies

We will study now the consequences of \eqref{<} and derive a
nonlinear PDE for $u_{1,0}$ which resemble the modified
dispersionless KP equation, and  was found in \cite{blaszak}. For
the sake of simplicity we write \eqref{<} as
\[
\partial_n\psi_<\cdot\psi_<^{-1}+P_<L^{n+1-r}=0.
\]

The right derivatives of $\psi_<$, as follows from
\eqref{right-derivative}, are
\[
\partial_n\psi_<\cdot\psi_<^{-1}=\partial_n\Psi_<+\frac{1}{2}\{\Psi_<,\partial_n\Psi_<\}
+\frac{1}{6}\{\Psi_<,\{\Psi_<,\partial_n\Psi_<\}\}+\cdots,
\]
so that
\begin{equation}\label{right-derivative-<}
\partial_n\psi_<\cdot\psi_<^{-1}=(\partial_n\Psi_1)p^{-r}+\big(\partial_n\Psi_2+
\frac{r}{2}(\Psi_{1,x}\partial_n\Psi_1-\Psi_1\partial_n\Psi_{1,x})\big)p^{-r-1}+{O}(p^{-r-2})
\end{equation}
for $p\to\infty$. Recall that when $p\to\infty$ we have
\[
L^{n+1-r}=p^{n+1-r}+u_{n+1-r,0}p^{n-r}+u_{n+1-r,1}p^{n-r-1}+u_{n+1-r,2}p^{n-r-2}+{O}(p^{n-r-3})
\]
with $u_{n+1-r,j}=-(n+1-r)\Psi_{j+1,x}+U_{n+1-r,j}$ and
$U_{n+1-r,j}$ a given nonlinear function of
$\Psi_1$,\dots,$\Psi_{j}$ and its $x$-derivatives. Equation
\eqref{<} together with \eqref{right-derivative-<} gives an
infinite set of equations, among which the two first  are
\begin{align*}
\partial_n\Psi_1&=-(n+1-r)\Psi_{n+1,x}+U_{n+1-r,n},\\
\partial_n\Psi_2+
\frac{r}{2}(\Psi_{1,x}\partial_n\Psi_1-\Psi_1\partial_n\Psi_{1,x})&=-(n+1-r)\Psi_{n+2,x}+U_{n+1-r,n+1}
\end{align*}
Thus, we get for $\Psi_{n+j,x}$, $j=1,2,\dots$  expressions in
terms of $\Psi_1,\dots,\Psi_{n}$ together with its $x$-derivatives
and integrals and also its $\partial_n$-derivative. For the next
flow we have
\begin{align*}
\partial_{n+1}\Psi_1&=-(n+2-r)\Psi_{n+2,x}+U_{n+2-r,n+1},\\
\partial_{n+1}\Psi_2+
\frac{r}{2}(\Psi_{1,x}\partial_{n+1}\Psi_1-\Psi_1\partial_{n+1}\Psi_{1,x})&=-(n+2-r)\Psi_{n+3,x}+U_{n+3-r,n+2}
\end{align*}
from where it follows a nonlinear PDE system for
$(\Psi_1,\dots,\Psi_n)$, in the variables $x,t_n,t_{n+1}$.

In particular, if $r\neq 2$ ---when $r=2$ the $t_1$-flow is
trivial--- and $n=1,2$ we get
\begin{align}
\label{flujo1}\Psi_{2,x}&=\frac{1}{2-r}\partial_1\Psi_1
+\frac{1}{2}\big(r\Psi_1\Psi_{1,xx}+(1-r)\Psi_{1,x}^2\big),\\
\label{flujo2}\partial_2\Psi_1&=\frac{3-r}{2-r}\Big(\partial_1\Psi_2+\frac{r}{2}\big(\Psi_{1,x}\partial_1\Psi_1-
\Psi_1\partial_1\Psi_{1,x}\big)\Big)\\&\quad+
(3-r)\Big(-\Psi_{1,x}\Psi_{2,x}+\frac{r}{2}\Psi_1\Psi_{1,x}\Psi_{1,xx}-\frac{r-1}{3}\Psi_{1,x}^3\Big),\notag
\end{align}
and hence,
\begin{multline}\label{mKP}
\Psi_{1,xt_2}=\frac{3-r}{(2-r)^2}\Psi_{1,t_1t_1}-\frac{(3-r)(1-r)}{2-r}\Psi_{1,xx}\Psi_{1,t_1}\\-
\frac{(3-r)r}{2-r}\Psi_{1,x}\Psi_{1,xt_1}-\frac{(3-r)(1-r)}{2}\Psi_{1,x}^2\Psi_{1,xx}.
\end{multline}

If we introduce
\begin{equation}\label{Psiu}
u:=u_{1,0}=-\Psi_{1,x},\quad \partial_x^{-1}u=\int_{x_0}^x
u(x)\mathrm{d}x
\end{equation}
we get for $u$ the following nonlinear PDE,
\begin{multline}\label{r-dmKP}
u_{t_2}=\frac{3-r}{(2-r)^2}(\partial_x^{-1}u)_{t_1t_1}+
\frac{(3-r)(1-r)}{2-r}u_x(\partial_x^{-1}u)_{t_1}+\frac{r(3-r)}{2-r}uu_{t_1}-
\frac{(3-r)(1-r)}{2}u^2u_x.
\end{multline}
This equation is similar to the dispersionless modified KP
equation which is recovered for $r=0$, hence we called it $r$-th
dispersionless modified KP ($r$-dmKP) equation, note that it was
derived for the first time in \cite{blaszak}. Therefore we will
refer to equation \eqref{mKP} as the potential $r$-dmKP equation.

\paragraph{The $\bar t_n$ flows for $\psi_>$}
From the intertwining property we find out that the $\bar
t_n$-flows for $\bar\Psi_k$ are derived from the $t_n$-flows for
$\Psi_k$ by replacing $r$ by $2-r$, and each $\partial_x^j\Psi_k$
by $(-1)^j\partial_x^j\bar\Psi_k$ so that \eqref{mKP} goes to
\begin{multline*}
-\bar\Psi_{1,x\bar t_2}=\frac{1+r}{r^2}\bar\Psi_{1,\bar t_1\bar
t_1}+\frac{(1+r)(1-r)}{r}\bar\Psi_{1,xx}\bar\Psi_{1,\bar t_1}\\-
\frac{(2-r)(1+r)}{r}\bar\Psi_{1,x}\bar\Psi_{1,x\bar
t_1}+\frac{(1+r)(1-r)}{2}\bar\Psi_{1,x}^2\bar\Psi_{1,xx}
\end{multline*}
or, in terms of $\bar v:=\bar v_1=-\bar\Psi_{1,x}$,
\begin{multline*}
-\bar v_{\bar t_2}=\frac{1+r}{r^2} \partial_x^{-1}\bar v_{\bar
t_1\bar t_1}-\frac{(1+r)(1-r)}{r}\bar v_{x}\partial_x^{-1}\bar
v_{\bar t_1}+
 \frac{(2-r)(1+r)}{r}\bar v\bar v_{\bar
t_1} +\frac{(1+r)(1-r)}{2}\bar v^2\bar v_{x}.
\end{multline*}

\subsection{The $r$-th dispersionless Dym hierarchy}

Here we shall discuss the consequences of the equations
\eqref{1-r} and \eqref{bar1-r}. In the one hand, we may rewrite
\eqref{1-r} as
\[
P_{1-r}L^{n+1-r}=\partial_n\psi_{1-r}\cdot\psi_{1-r}^{-1},
\]
which in terms of $\bar\psi_{1-r}:=\psi_{1-r}^{-1}=\exp(-\xi
p^{1-r})$ reads as
\begin{equation}\label{1-ru}
\partial_n\bar\psi_{1-r}\cdot\bar\psi_{1-r}^{-1}+\Ad_{\bar\psi_{1-r}} (u_{n+1-r,n-1}
p^{1-r})=0.
\end{equation}
On the other hand \eqref{bar1-r} reads
\begin{equation}\label{1-rv}
\bar\partial_n\psi_{1-r}\cdot\psi_{1-r}^{-1}+ \Ad_{\psi_{1-r}}
(\bar v_{1-r-n,n-1}p^{1-r})=0.
\end{equation}

At this point is useful to  recall \eqref{Xpsi} which reads
\[
\Ad_{\exp(\xi
p^{1-r})}(f(x)p^n)=\left\{\begin{aligned}&X_x(x)^{-\frac{n}{1-r}}f(X(x))
p^n,& \int_x^X\frac{\d x}{\xi (x)}&=(1-r), && r\neq 1,\\
&\exp(-n \xi_x(x))f(x)p^n,&& & &r=1,
\end{aligned}\right.
\]
where have used
\[
X_x=\frac{\xi(X(x))}{\xi(x)}.
\]

In particular \eqref{Xpsi} implies for $r\neq 1$ the following
equations
\[
\begin{aligned}
\Ad_{\exp(\xi p^{1-r})}(x)&=X,&\int_x^{X}\frac{\d x}{\xi
(x)}=(1-r),\\
\Ad_{\exp(-\xi p^{1-r})}(x)&=\bar X,&\int_x^{\bar X}\frac{\d
x}{\xi (x)}=-(1-r),
\end{aligned}
\]

Observe that $\bar X$ is the inverse function of $X$; i.e., $\bar
X\circ X=\text{id}$. This is also a consequence $x=\Ad_{\exp(-\xi
p^{1-r})}(X(x))=\bar X(X(x))$. Another remarkable fact is that $X$
is the canonical transform of the $x$ variable under $\psi_{1-r}$.
Note also that the conjugate momenta to the variables $X$ and
$\bar X$ are
\[
\Ad_{\exp(\xi p^{1-r})}(p)=(X_x)^{-\frac{1}{1-r}}p,\quad
\Ad_{\exp(-\xi p^{1-r})}(p)=(\bar X_x)^{-\frac{1}{1-r}}p,
\]
respectively. Finally, notice that when $r=1$ we have $X(x)=\bar
X(x)=x$.

Now, if $\partial$ is a given Lie algebra derivation then
\begin{equation}\label{beta}\begin{aligned}
\partial\psi_{1-r}\cdot\psi_{1-r}^{-1}&=\beta(x) p^{1-r},\\
\partial\bar\psi_{1-r}\cdot\bar\psi_{1-r}^{-1}&=\bar\beta(x)
p^{1-r},
\end{aligned}
\end{equation}
but
\[
\partial\psi_{1-r}\cdot\psi_{1-r}^{-1}+\Ad_{\psi_{1-r}}(\partial\bar\psi_{1-r}\cdot\bar\psi_{1-r}^{-1})=0\Rightarrow
\beta(x) p^{1-r}+\Ad_{\psi_{1-r}}(\bar\beta(x) p^{1-r})=0,
\]
so that
\[
\beta(x)+\bar\beta(X(x))X_x(x)^{-1}=0,
\]
or, upon the use of $X_x(\bar X(x))\bar X_x=1$,
\begin{equation}\label{beta2}
\beta(\bar X(x))+\bar\beta(x)\bar X_x(x)=0.
\end{equation}
 From \eqref{beta} we get
\[
\begin{aligned}
\partial X&=\{\beta(x) p^{1-r}, X(x)\}=(1-r)\beta X_x,\\
\partial \bar X&=\{\bar\beta(x) p^{1-r},\bar X(x)\}=(1-r)\bar\beta\bar
X_x,
\end{aligned}
\]
so that, when $r\neq 1$,
\[
\beta=\frac{1}{1-r}\frac{\partial X}{X_x},\quad
\bar\beta=\frac{1}{1-r}\frac{\partial \bar X}{\bar X_x}.
\]
Observe that the compatibility with \eqref{beta2} follows from
that $\partial X(\bar X(x))+X_x(\bar X)\partial \bar X(\bar
X(x))=0$.

Therefore, we have proven that
\[
\begin{aligned}
\partial\psi_{1-r}\cdot\psi_{1-r}^{-1}&=\left\{\begin{aligned}&\dfrac{1}{1-r}\dfrac{\partial
X}{X_x}p^{1-r},& \int_x^{X}\frac{\d x}{\xi (x)}&=(1-r),& &r\neq 0,
\\ &\partial\xi\,p^{1-r},& && &r=1,
\end{aligned}\right.\\
\partial\bar\psi_{1-r}\cdot\bar\psi_{1-r}^{-1}&=\left\{\begin{aligned}&\dfrac{1}{1-r}\dfrac{\partial
\bar X}{\bar X_x}p^{1-r},& \int_x^{\bar X}\frac{\d x}{\xi
(x)}&=-(1-r),&&r\neq 0,
\\& -\partial\xi\,p^{1-r},&&& & r=1.
\end{aligned}\right.
\end{aligned}
\]

We are now ready to tackle \eqref{1-ru} and \eqref{1-rv}, which
read
\begin{align}
\label{n1-ru}&\left\{\begin{aligned}\partial_n \bar X&=-(1-r)u_{n+1-r,n-1}(\bar X),& r&\neq 1,\\
\partial_n\xi&=u_{n,n-1}, &r&=1,\end{aligned}\right.\\
\label{barn1-rv}&\left\{\begin{aligned} \bar\partial_n X
&=-(1-r)\bar
v_{1-r-n,n-1}(X),& r&\neq 1,\\
\bar\partial_n\xi&=-\bar v_{-n,n-1},& r&=1.
\end{aligned}\right.
\end{align}

Let us look at the consequences of \eqref{n1-ru} for $n=1$ and
$n=2$ and recall \eqref{u}:
\begin{align}
\label{flujo11-r}&\left\{\begin{aligned}\bar
X_{t_1}&=(1-r)(2-r)\Psi_{1,x}(\bar
X),& r &\neq 1\\
\xi_{t_1}&=-\Psi_{1,x},&r &=1,\end{aligned}\right.\\
\label{flujo21-r}&\left\{\begin{aligned}
 & \begin{aligned} \bar
X_{t_2}=-(1-r)(3-r)&\Big(\Psi_{2,x}(\bar
X)\\&-\frac{1}{2}(r\Psi_1(\bar X)\Psi_{1,xx}(\bar X)+
(2-r)\Psi_{1,x}(\bar X)^2)\Big),
\end{aligned}
 &r&\neq 1,\\
&\xi_{t_2}=-2(\Psi_{2,x}-\frac{1}{2}(\Psi_1\Psi_{1,xx}+\Psi_{1,x}^2)),
&r&=1.
\end{aligned}\right.
\end{align}

We first analyze the case $r\neq 1$. The first equation
\eqref{flujo11-r}, when $r\neq 1,2$, gives the important relation
\begin{equation}\label{psiX}
\Psi_{1,x}(\bar X)=\frac{1}{(1-r)(2-r)}\bar X_{t_1}.
\end{equation}
By introducing \eqref{flujo1} into \eqref{flujo21-r} we get
\[
\bar X_{t_2}=(1-r)(3-r)\Big(\frac{1}{2-r}\Psi_{1,t_1}(\bar
X)-\frac{1}{2}\Psi_{1,x}(\bar X)^2\Big),
\]

which we are going to manipulate. Firstly, we take its $x$
derivative
\[
\bar X_{xt_2}=(1-r)(3-r)\Big(\frac{1}{2-r}\Psi_{1,t_1x}(\bar
X)\bar X_x-\Psi_{1,x}(\bar X)(\Psi_{1,x}(\bar X))_x\Big),
\]
second we see that
\[
\begin{aligned}
\Psi_{1,t_1x}(\bar X)\bar X_x=&\big((\Psi_{1,x}(\bar
X))_{t_1}-\Psi_{1,xx}(\bar X)\bar X_{t_1}\big)\bar
X_{x}\\=&(\Psi_{1,x}(\bar X))_{t_1}\bar X_x-(\Psi_{1,x}(\bar
X))_x\bar X_{t_1}.
\end{aligned}
\]
Therefore,
\begin{multline*}
\bar X_{xt_2}=(1-r)(3-r)\Big(\frac{1}{2-r}\big((\Psi_{1,x}(\bar
X))_{t_1}\bar X_x-(\Psi_{1,x}(\bar X))_x\bar
X_{t_1}\big)-\Psi_{1,x}(\bar X)(\Psi_{1,x}(\bar X))_x\Big),
\end{multline*}
that recalling \eqref{psiX} reads as follows
\begin{equation}\label{PDEbarX}
\bar X_{xt_2}=\frac{3-r}{2-r}\Big(\frac{1}{2-r}\bar X_{t_1t_1}\bar
X_x-\frac{1}{1-r}\bar X_{xt_1}\bar X_{t_1}\Big)
\end{equation}

In the case $r=1$ we  introduce \eqref{flujo1} into
\eqref{flujo21-r} to get
\[
\xi_{t_2}=-2\Psi_{1,t_1}+\Psi_{1,x}^2,
\]
that taking the $x$-derivative reads
\[
\xi_{xt_2}=-2\Psi_{1,xt_1}+2\Psi_{1,x}\Psi_{1,xx},
\]
and recalling \eqref{flujo11-r} for $r=1$ is
\begin{equation}\label{PDExi}
\xi_{t_2x}-2\xi_{t_1}\xi_{t_1x}-2\xi_{t_1t_1}=0,
\end{equation}

Upon the introduction of the variable
\begin{equation}\label{vX}
v=\begin{cases}(\bar X_x)^{-\frac{1}{1-r}},& r\neq 1,\\
\exp\xi_x, & r=1,
\end{cases}
\end{equation}
equations \eqref{PDEbarX} and \eqref{PDExi} transforms onto
\begin{equation}\label{r-dDym}
v_{t_2}=\frac{3-r}{(2-r)^2}v^{r-1}\Big(v^{2-r}\partial_x^{-1}(v^{r-2}v_{t_1})\Big)_{t_1}
\end{equation}
which resembles the dispersionless Dym  equation which appears for
$r=0$, hence we refer to it as the $r$-th dispersionless Dym
($r$-dDym) equation (for $r\neq 2$); this equation was first
derived in \cite{blaszak}. We shall refer to \eqref{PDEbarX} as
the potential $r$-dDym equation.

From \eqref{barn1-rv} we derive
\begin{align} \label{barflujo11-r}
&\left\{\begin{aligned}X_{\bar t_1}&=-(1-r) r\bar\Psi_{1,x}(X),&r&\neq 1\\
\xi_{\bar t_1}&=-\bar\Psi_{1,x},& r&=1\end{aligned}\right.\\
\label{barflujo21-r} &\left\{\begin{aligned}
&\begin{aligned}X_{\bar
t_2}=-(1-r)(1+r)&\Big(\bar\Psi_{2,x}(X)\\&+\frac{1}{2}((2-r)\bar\Psi_1(X)\bar\Psi_{1,xx}(
X)+ r\bar\Psi_{1,x}(X)^2)\Big)
\end{aligned},& r&\neq 1,
\\
&\xi_{\bar
t_2}=-2(\bar\Psi_{2,x}+\frac{1}{2}(\bar\Psi_1\bar\Psi_{1,xx}+\bar\Psi_{1,x}^2)),&
r&=1,\end{aligned}\right.
\end{align}
which, for $r\neq 1$, lead to
\begin{equation}\label{PDEX}
-X_{x\bar t_2}=\frac{1+r}{r}\Big(\frac{1}{r}X_{\bar t_1\bar t_1}
X_x+\frac{1}{1-r}X_{x\bar t_1}X_{\bar t_1}\Big),
\end{equation}
and when $r=1$ to
\begin{equation}\label{barPDExi}
\xi_{\bar t_2x}+2\xi_{\bar t_1}\xi_{\bar t_1x}+2\xi_{\bar t_1\bar
t_1}=0.
\end{equation}

Observe that equations \eqref{barn1-rv} can be obtained from
\eqref{n1-ru} with the use of the intertwining \eqref{intertwin}
in the following two steps:
 \begin{enumerate}
 \item We have
\[
\partial_n'\bar X'=-(1-r')u_{n'+1-r',n'-1}'(\bar X'),
\]
that recalling that $\partial_n'=\bar\partial_n$, $r'=2-r$ and
$u_{n'+1-r',n'-1}'(\bar X')=\bar v_{1-r-n,n-1}(-\bar X')$
($x'=-x$) reads
\[
\bar\partial_n\bar X'=(1-r)\bar v_{1-r-n,n-1}(-\bar X').
\]
\item Now, we only need to find $\bar X'$; this can be done in at
least two ways. From the definition we have
\[
\bar X'=\Ad_{{\psi^{'-1}_{1-r'}}}(x')=\Ad_{\psi_{1-r}}(-x)=-X,
\]
an alternative is from
\[
\int_{x'}^{\bar X'}\frac{\d x}{\xi'(x)}=-(1-r')\Rightarrow
\int_{-x}^{\bar X'}\frac{\d x}{-\xi(-x)}=(1-r)\Rightarrow
\int_{x}^{-\bar X'}\frac{\d x}{\xi(x)}=(1-r).
\]
So that $\bar X'=-X$. \item For $r=1$ we only need to recall that
$\psi^{'-1}_{0}=\psi_0$ so that $\xi'(x')=-\xi(-x)$.
\end{enumerate}

\subsection{Miura map among $r$-dmKP and $r$-dDym equations}

From \eqref{Psiu}, \eqref{psiX} and \eqref{vX} we  get
\[
\left\{\begin{split} -\frac{1}{(1-r)(2-r)}\bar X_{t_1}&=u(\bar
X),& \bar
X&=\partial_x^{-1}v^{r-1},& r&\neq 1,\\
\xi_{t_1}&=u,& \xi&=\partial_x^{-1}\log v,&  r&=1
\end{split}\right.
\]
and
\begin{equation}\label{miura1}
\left\{ \begin{split}
\partial_x^{-1}(v^{r-1})_{t_1}&=-(2-r)(1-r)u(\partial_x^{-1}v^{r-1}),&r&\neq
1\\
\partial_x^{-1}(\log v)_{t_1}&=u(x),& r&=1.
\end{split}
\right.
\end{equation}
Equations \eqref{miura1} relate solutions $u$ and $v$ of the
$r$-dmKP \eqref{r-dmKP} and $r$-dDym \eqref{r-dDym} equations.

If we derive with respect to $x$ the above relations we get
\begin{equation}\label{miura}
\left\{\begin{split}u_x(\partial_x^{-1}v^{r-1})&=\frac{1}{2-r}(\log
v)_{t_1},& r&\neq 1,\\
u_x&=(\log v)_{t_1},& r&=1.
\end{split}\right.
\end{equation}
 In any case observe that a solution $v$ to the $r$-dDym equation
provide us with a solution of the $r$-dmKP equation, after the
calculation of some inverse functions of $\bar X$, but the
reverse, given  a solution $u$ of the $r$-dmKP equation
\eqref{r-dmKP} to get $v$ a solution of the $r$-dDym equation
\eqref{r-dDym} do not follow  from formulae either \eqref{miura1}
neither \eqref{miura}. Observe that in \cite{chen-tu} a similar
Miura map was derived, in a quite different manner, for the well
known $r=0$ case.

\subsection{The $r$-th dispersionless Toda hierarchy}

Here we consider  equations \eqref{>} and \eqref{bar<}
\begin{align*}
\Ad_{\bar\psi_{1-r}}P_>L^{n+1-r}&=\partial_n\psi_>\cdot\psi_>^{-1},\\
\Ad_{\psi_{1-r}}P_<\bar
\ell^{1-r-n}&=\bar\partial_n\psi_<\cdot\psi_<^{-1},
\end{align*}
that for $n=1$ reads
\begin{align}
\label{toda11}\Ad_{\bar\psi_{1-r}}P_>L^{2-r}&=\partial_1\psi_>\cdot\psi_>^{-1},\\
\label{bartoda1}\Ad_{\psi_{1-r}}P_<\bar
\ell^{-r}&=\bar\partial_1\psi_<\cdot\psi_<^{-1}.
\end{align}
Looking at the leading terms in $p$ we obtain from \eqref{toda11}
and \eqref{bartoda1} the following equations
\begin{align*}\bar
\Psi_{1,t_1}&=
\left\{\begin{aligned}(\bar X_x)^{-\frac{2-r}{1-r}},& &r\neq 1,\\
\exp(\xi_x),& &r= 1,
\end{aligned}\right.\\
\Psi_{1,\bar t_1}&=\left\{\begin{aligned}(X_x)^{\frac{r}{1-r}},& &
r\neq 1,\\
\exp(\xi_x),& & r= 1.\end{aligned}\right.
\end{align*}
Recall now equations \eqref{psiX} and \eqref{barflujo11-r} which,
taking into account $\bar X_{t_1}(X)=-X_{t_1}(x)/X_x(x)$ and $
X_{\bar t_1}(\bar X)=-\bar X_{\bar t_1}(x)/\bar X_x(x)$, can be
written as
\begin{align*}
\Psi_{1,x}&=\left\{\begin{aligned}&-\frac{1}{(1-r)(2-r)}\frac{X_{t_1}}{X_x},& r&\neq 1,\\
&-\xi_{t_1},&  r&=1,\end{aligned}\right.\\
\bar \Psi_{1,x}&=\left\{\begin{aligned}&\frac{1}{(1-r)r}\frac{\bar
X_{\bar t_1}}{\bar X_x},&& &r\neq 1,\\
&-\xi_{\bar t_1},& &&r=1,\end{aligned}\right.
\end{align*}
respectively. The compatibility of these equations lead to
%\begin{align}
% &\left\{\begin{aligned}\Big((\bar
%X_x)^{-\frac{2-r}{1-r}}\Big)_x-\frac{1}{(1-r)r}\Big(\frac{\bar
%X_{\bar t_1}}{\bar X_x}\Big)_{ t_1}=0,\label{toda}\\
%\Big((X_x)^{\frac{r}{1-r}}\Big)_x+\frac{1}{(1-r)(2-r)}\Big(\frac{
%X_{t_1}}{ X_x}\Big)_{\bar t_1}=0,\end{aligned}\right.& &r\neq 1,\\
%&(\exp(\xi_x))_x+\xi_{t_1\bar t_1}=0, && r=1,\\
%\label{bartoda}\end{align}

\begin{subequations}
\begin{align} \Big((\bar
X_x)^{-\frac{2-r}{1-r}}\Big)_x-\frac{1}{(1-r)r}\Big(\frac{\bar
X_{\bar t_1}}{\bar X_x}\Big)_{ t_1}=0,\label{toda}\\
\Big((X_x)^{\frac{r}{1-r}}\Big)_x+\frac{1}{(1-r)(2-r)}\Big(\frac{
X_{t_1}}{ X_x}\Big)_{\bar t_1}=0,\label{toda1}
\end{align} when $ r\neq 1$ while for $r=1$ the
\end{subequations} equation is
\begin{equation}\label{toda0}
(\exp(\xi_x))_x+\xi_{t_1\bar t_1}=0,
\end{equation}
 this are new integrable equations,
which we call  $r$-th dispersionles Toda ($r$-dToda) equation,
because for $r=1$ the corresponding equation is the dispersionless
Toda equation
---known also as the Boyer--Finley equation---.

Equations \eqref{toda} and \eqref{toda1} are the same equation,
indeed. To prove it we just need to evaluate equation
\eqref{toda1} on $\bar X$ and recall that $X$ is the inverse
function of  $\bar X$, $X(\bar X(x))=x$, so that
\[
\begin{aligned}
X_x(\bar X(x))\bar X_x(x)&=1,\\
X_{t_1}(\bar X(x))+X_x(\bar X(x))\bar X_{t_1}(x)&=0.
\end{aligned}
\]
From the relations
\[
\begin{split}
\frac{X_{t_1}(\bar X(x))}{X_x(\bar X(x))}&=-\bar X_{t_1},\\
\Big(\frac{X_{t_1}}{X_x}\Big)_x(\bar X)&=
\Big(\frac{X_{t_1}(\bar X)}{X_x(\bar X)}\Big)_{x}\frac{1}{\bar X_x},\\
\Big(\frac{X_{t_1}}{X_x}\Big)_{\bar t_1}(\bar
X)&=\Big(\frac{X_{t_1}(\bar X)}{X_x(\bar X)}\Big)_{\bar t_1}-
\Big(\frac{X_{t_1}}{X_x}\Big)_{x}(\bar X)\bar X_{\bar t_1}
\end{split}
\]
we derive
\[
\Big(\frac{X_{t_1}}{X_x}\Big)_{\bar t_1}(\bar X)=-\bar X_{t_1\bar
 t_1}+\frac{\bar X_{xt_1}\bar X_{\bar t_1}}{\bar X_x}=-\bar X_x
 \Big(\frac{\bar X_{\bar t_1}}{\bar X_x}\Big)_{t_1}.
\]
We evaluate now
\[
\big((X_x)^{\frac{r}{1-r}}\big)_x(\bar X)=\big((X_x(\bar
X))^{\frac{r}{1-r}}\big)_x\frac{1}{\bar X_x}=\big((\bar
X_x)^{-\frac{r}{1-r}}\big)_x\frac{1}{\bar X_x}.
\]
Therefore, we conclude that \eqref{toda1} imply
\[
\big((\bar X_x)^{-\frac{r}{1-r}}\big)_x\frac{1}{{\bar
X_x}^2}-\frac{1}{(1-r)(2-r)} \Big(\frac{\bar X_{\bar t_1}}{\bar
X_x}\Big)_{t_1}=0
\]
but observing
\[
\big((\bar X_x)^{-\frac{r}{1-r}}\big)_x\frac{1}{{\bar
X_x}^2}=\frac{r}{2-r}\big((\bar X_x)^{-\frac{2-r}{1-r}}\big)_x
\]
we deduce, as claimed, equation \eqref{toda}.
%and we conclude
%\[
%\frac{X_{t_1}(\bar X(x))}{X_x(\bar X(x))}=-\bar X_{t_1}(x)
%\]

\section{Additional symmetries}

In this section we deal with the additional symmetries of the
integrable hierarchies just described. We first introduced the
Orlov functions  $M, \bar M, \bar m$ in this context and the
consider the construction of additional symmetries. We compute
explicitly some of these additional symmetries for the potential
$r$-dmKP \eqref{mKP}, the $r$-dDym \eqref{PDEbarX} and the
$r$-dToda \eqref{toda} equations, finding explicit symmetries of
these nonlinear equations depending on arbitrary functions of the
variable $t_2$.

\subsection{The Orlov funtions}

In formulae  \eqref{lax_def} we introduced the Lax functions $L,
\bar\ell$ and $\bar L$, which are the canonical transformation of
the $p$ variable through $\psi_<,\psi_>$ and $\psi_{\geqslant}$,
respectively. Recalling that $t$ and $\bar t$ are functions of $p$
only we can write these Lax functions as follows:
\begin{align*}
L&=\Ad_{\psi_<\cdot\exp t}p,&\bar \ell&=\Ad_{\psi_>\cdot\exp \bar
t}p,&\bar L&=\Ad_{\psi_\geqslant\cdot\exp \bar t}p.
\end{align*}
The Orlov functions $M, \bar m$ and $\bar M$ are defined
analogously with the replacement of $p$ by $x$:
\begin{equation}\label{orlov}
\begin{aligned}
M&:=\Ad_{\psi_<\cdot\exp t}x,&\bar m&:=\Ad_{\psi_>\cdot\exp \bar
t}x,&\bar M&:=\Ad_{\psi_\geqslant\cdot\exp \bar t}x.
\end{aligned}
\end{equation}

In the next Proposition we describe the form of the Orlov
functions as series in the Lax functions.

\begin{pro}\label{pro-orlov}
The Orlov functions defined in \eqref{orlov} have the following
expansions
\begin{align*}
M&=\cdots+w_2L^{-2}+w_1L^{-1}+x+(2-r)t_1L+(3-r)t_2L^2+\cdots,\quad
L\to\infty
\\
\bar m&=\cdots-(r+1)\bar t_2\bar \ell^{-2}-r\bar t_1\bar
\ell^{-1}+x+\bar \omega_1\bar \ell +\bar \omega_2\bar
\ell^2+\cdots,\quad\bar \ell\to 0
\\
\bar M &=\cdots-(r+1)\bar t_2\bar L^{-2}-r\bar t_1\bar
L^{-1}+X+\bar w_1\bar L +\bar w_2\bar L^2+\cdots,\quad \bar L\to 0
\end{align*}
with
\begin{align*}
&\begin{aligned}
w_1&=-r\Psi_1,\\w_2&=-(r+1)\big(\Psi_2-\frac{1}{2}
r\Psi_1\Psi_{1,x}\big),\\ \vdots&
\end{aligned}\\
&\begin{aligned}
 \bar \omega_1&=(2-r)\bar\Psi_1(x),\\
  \bar \omega_2&=(3-r)\big(\bar\Psi_2(x)-\frac{1}{2}
(2-r)\bar\Psi_1(x)\bar\Psi_{1,x}(x)\big),\\ \vdots&
\end{aligned}
\end{align*}
and \begin{align*} &\begin{aligned}
 \bar w_1&=(2-r)\bar\Psi_1(X),\\
  \bar w_2&=(3-r)\big(\bar\Psi_2(X)-\frac{1}{2}
(2-r)\bar\Psi_1(X)\bar\Psi_{1,x}(X)\big),\\ \vdots&
\end{aligned}
\end{align*}
\end{pro}

\begin{proof}
Now, taking into account that
\[
\ad_t x=\{t,x\}=p^r\frac{\partial t}{\partial
p}=(2-r)t_1p+(3-r)t_2p^2+\cdots
\]
 we evaluate
\[
\Ad_{\exp t}x=\exp(\ad_t)(x)=x+p^r\frac{\partial t}{\partial
p}=x+(2-r)t_1p+(3-r)t_2p^2+\cdots,
\]
and, therefore,
\[
\begin{aligned}
M&=\Ad_{\psi_<}\big(x+p^r\frac{\partial t}{\partial
p}\big)=\Ad_{\psi_<}(x)+L^r\frac{\partial t(L)}{\partial
L}\\&=\Ad_{\psi_<}(x)+(2-r)t_1L+(3-r)t_2L^2+\cdots.
\end{aligned}
\]
To compute $M$ we need to evaluate
\[
\ad_{\Psi_<}(x)=\{\Psi_<,x\}=p^r\frac{\partial \Psi_<}{\partial
p}=D_p\Psi_<
\]
where
\[
D_p:=p^r\frac{\partial}{\partial p}.
\]
Notice that $D_p$ is a derivation of the Lie algebra $\g$:
\[
D_p\{f,g\}=\{D_pf,g\}+\{f,D_pg\}.
\]
Thus
\[
\Ad_{\psi_<}
(x)=x+\sum_{n=0}^{\infty}\frac{1}{(n+1)!}\ad_{\Psi_<}^nD_p\Psi_<=x+D_p\psi_<\cdot\psi_<^{-1},
\]
where
\[
D_p\Psi_<=-(r\Psi_1p^{-1}+(r+1)\Psi_2p^{-2}+\cdots).
\]
We can compute now $D_p\psi_<\cdot\psi_<^{-1}$:
\begin{align*}
D_p\psi_<\cdot\psi_<^{-1}&=D_p\Psi_<+\frac{1}{2}\{\Psi_<,D_p\Psi_<\}+\cdots\\&=
-r\Psi_1p^{-1}-\big((r+1)\Psi_2-\frac{1}{2}r(r-1)\Psi_1\Psi_{1,x}\big)p^{-2}+\cdots.
%\big(-(r+2)\Psi_3+\Psi_{1,x}\Psi_2-\frac{1}{2}\Psi_1\Psi_{2,x}\big)p^{-3}+\cdots,
\end{align*}

From
\[
L^m=p^m+u_{m,0}p^{m-1}+u_{m,1}p^{m-2}+\cdots
\]
we deduce
\[
p^m=L^m-u_{m,0}L^{m-1}-(u_{m,1}-u_{m-1,1}u_{m,0})L^{m-2}-\cdots
\]
so that
\begin{equation}\label{MinL}
M=\cdots+w_2L^{-2}+w_1L^{-1}+x+(2-r)t_1L+(3-r)t_2L^2+\cdots,\\
\end{equation}
where, for example
\begin{align*}
w_1&=-r\Psi_1,\\w_2&=-(r+1)\big(\Psi_2-\frac{1}{2}
r\Psi_1\Psi_{1,x}\big).
\end{align*}

For $\bar m$ and $\bar M$ we proceed in a similar manner. First
\[
\begin{aligned}
\ad_{\bar t} x&=\{\bar t,x\}=p^r\frac{\partial \bar t}{\partial
p}=-r\bar t_1p^{-1}-(r+1)\bar t_2p^{-2}+\cdots\,\\ \Ad_{\exp \bar
t}x&=x+p^r\frac{\partial \bar t}{\partial p}=x-r\bar
t_1p^{-1}-(r+1)\bar t_2 p^{-2}+\cdots,
\end{aligned}
\]
so that
\[
\begin{aligned}
\bar m&=\Ad_{\psi_>}(x)+\bar \ell^r\frac{\partial \bar
t(\bar\ell)}{\partial \bar\ell}\\&=\Ad_{\psi_>}(x)-r\bar
t_1\bar\ell^{-1}-(r+1)\bar t_2\bar \ell^{-2}+\cdots,\\ \bar
M&=\Ad_{\psi_\geqslant}(x)+\bar L^r\frac{\partial t(\bar
L)}{\partial\bar L}\\&=\Ad_{\psi_\geqslant}(x)-r\bar t_1\bar
L^{-1}-(r+1)\bar t_2\bar L^{-2}+\cdots.
\end{aligned}
\]
Now
\[ \Ad_{\psi_>} (x)=x+D_p\psi_>\cdot\psi_>^{-1}.
\]
and
\[
D_p\Psi_>=(2-r)\bar\Psi_1p+(3-r)\bar\Psi_2p^{2}+\cdots.
\]
Hence,
\begin{align*}
D_p\psi_>\cdot\psi_>^{-1}&=D_p\bar\Psi+\frac{1}{2}\{\bar\Psi,D_p\bar\Psi\}+\cdots\\&=
(2-r)\bar\Psi_1p+\Big((3-r)\bar\Psi_2+\frac{1}{2}(2-r)(1-r)\bar\Psi_1\bar\Psi_{1,x}\Big)p^{2}+\cdots,
\end{align*}

From
\[
\bar\ell^m=p^m+\bar v_{m,0}p^{m+1}+\bar v_{m,1}p^{m+2}+\cdots
\]
we deduce
\[
p^m=\bar\ell^m-\bar v_{m,0}\bar\ell^{m+1}-(\bar v_{m,1}-\bar
v_{m+1,0}\bar v_{m,0})\bar\ell^{m-2}-\cdots
\]
so that
\[
\bar m=\cdots+(3-r)\big(\bar\Psi_2-\frac{1}{2}
(2-r)\bar\Psi_1\bar\Psi_{1,x}\big)\bar\ell^2+(2-r)\bar\Psi_1(x)\bar\ell+x-r\bar
t_1\bar\ell^{-1}-(r+1)\bar t_2\bar\ell^{-2}+\cdots.
\]
Therefore,
\begin{align} \bar M&=\Ad_{\psi_{1-r}}\bar
m=\cdots-(r+1)\bar t_2\bar L^{-2}-r\bar t_1\bar L^{-1}+X+\bar
w_1\bar L +\bar w_2\bar L^2+\cdots.\label{barm}
\end{align}
with \begin{align*}
 \bar w_1&=(2-r)\bar\Psi_1(X),\\
  \bar w_2&=(3-r)\big(\bar\Psi_2(X)-\frac{1}{2}
(2-r)\bar\Psi_1(X)\bar\Psi_{1,x}(X)\big).
\end{align*}
\end{proof}
We now find the Lax equations for $M$ and $\bar M$. For that aim
we compute
\begin{align*}
&\partial_n(\psi_<\cdot\exp t)\cdot (\psi_<\cdot\exp
t)^{-1}=\partial\psi_<\cdot\psi_<^{-1}+\Ad_{\psi_<}(
p^{n+1-r})=P_\geqslant L^{n+1-r},\\
&\bar \partial_n(\psi_\geqslant\cdot\exp \bar t)\cdot
(\psi_\geqslant\cdot\exp \bar
t)^{-1}=\partial\psi_\geqslant\cdot\psi_\geqslant^{-1}+\Ad_{\psi_\geqslant}(
p^{-n+1-r})=P_<\bar L^{-n+1-r}
\end{align*}
and conclude that
\begin{pro}\label{Lax-Orlov}
The Lax and Orlov functions are subject to the following Lax
equations:
\begin{align*} &\left\{\begin{aligned}
\partial_n L&=\{P_\geqslant L^{n+1-r},L\},&
\partial_n\bar L&=\{P_{\geqslant}L^{n+1-r},\bar L\},\\
\bar\partial_n L&=\{P_<\bar  L^{1-r-n},L\},&
 \bar\partial_n\bar L&=\{P_< \bar L^{1-r-n},\bar L\},
\end{aligned}\right.\\
&\left\{\begin{aligned}
\partial_n M&=\{P_\geqslant L^{n+1-r},M\},&
\partial_n\bar M&=\{P_{\geqslant}L^{n+1-r},\bar M\},\\
\bar\partial_n M&=\{P_<\bar L^{1-r-n},M\},&
 \bar\partial_n\bar M&=\{P_<\bar L^{1-r-n},\bar M\}.
\end{aligned}\right.
\end{align*}
\end{pro}
\subsection{Additional symmetries and its generators}

Let us consider that the initial conditions $h,\bar h$ in the
factorization problem describe smooth curves $h(s)=\exp(H(s)),\bar
h(s)=\exp(\bar H(s))$ in $G$
---here $H(s)$ and $\bar H(s)$ are curves in $\g$---. This imply that the factors $\psi_<=\psi_<(s)$ and
$\psi_\geqslant=\psi_\geqslant(s)$ in the corresponding
factorization problemdo depend on $s$:
\begin{equation}\label{factorization-s}
\psi_<(s)\cdot \exp(t)\cdot h(s)=\psi_\geqslant(s)\cdot\exp(\bar
t)\cdot\bar h(s).
\end{equation}
If we introduce the notation
\begin{equation}\label{F}
\begin{aligned}
F&:=\frac{\d h}{\d s}\cdot h^{-1},\\
\bar F&:=\frac{\d \bar h}{\d s}\cdot \bar h^{-1}
\end{aligned}
\end{equation}
and take the right-derivative with respect to $s$  of
\eqref{factorization-s} we get
\[
\frac{\d\psi_<}{\d s}\cdot\psi_<^{-1}+\Ad_{\psi_<\cdot\exp
t}(F(p,x))= \frac{\d\psi_\geqslant}{\d
s}\cdot\psi_\geqslant^{-1}+\Ad_{\psi_\geqslant\cdot\exp\bar
t}(\bar F(p,x))
\]
that implies
\[
\frac{\d\psi_\geqslant}{\d
s}\cdot\psi_\geqslant^{-1}-\frac{\d\psi_<}{\d
s}\cdot\psi_<^{-1}=F(L,M)-\bar F(\bar L,\bar M).
\]
Now,  we may split this equation into
\begin{equation}\label{symmetries-fac}
\begin{aligned}
\frac{\d\psi_<}{\d s}\cdot\psi_<^{-1}&=-P_<(F(L,M)-\bar F(\bar
L,\bar M)),\\\frac{\d\psi_\geqslant}{\d
s}\cdot\psi_\geqslant^{-1}&=P_\geqslant(F(L,M)-\bar F(\bar L,\bar
M)).
\end{aligned}
\end{equation}
 Equations \eqref{symmetries-fac} imply for the Lax and Orlov
 functions $L,\bar L, M$ and $\bar M$ the
\begin{pro}\label{add_symm_Lax_Orlov}
The Lax and Orlov functions are transformed by the additional
symmetries according to the following formulae
 \begin{equation}\label{symmetries-laxorlov}
\begin{aligned}
\frac{\d L}{\d s}&=\{-P_<(F(L,M)-\bar F(\bar L,\bar M)),L\},
&\frac{\d M}{\d s}&=\{-P_<(F(L,M)-\bar F(\bar L,\bar M)),M\},\\
\frac{\d \bar L}{\d s}&=\{P_\geqslant(F(L,M)-\bar F(\bar L,\bar
M)),\bar L\}, &\frac{\d \bar M}{\d s}&=\{P_\geqslant(F(L,M)-\bar
F(\bar L,\bar M)),\bar M\}. \end{aligned}
 \end{equation}
\end{pro}
As we know there is no loss of generality if we take $\bar
h(s)=\text{id}$, then $\bar F=0$ and \eqref{symmetries-fac} read
as
 %\begin{subequations}%\begin{equation}
\begin{align}\label{symmetries<}
\frac{\d \psi_<}{\d s}\cdot\psi_<^{-1}&=-P_<(F(L,M)),\\
\label{symmetries>}\frac{\d \psi_\geqslant}{\d
s}\cdot\psi_\geqslant^{-1}&=P_\geqslant(F(L,M)).
\end{align}

Alternatively, we could set $h(s)=\text{id}$ so that
\begin{align}\label{bar-symmetries<}
\frac{\d \psi_<}{\d s}\cdot\psi_<^{-1}&=P_<(\bar F(\bar L,\bar M)),\\
\label{bar-symmetries>}\frac{\d \psi_\geqslant}{\d
s}\cdot\psi_\geqslant^{-1}&= -P_\geqslant(\bar F(\bar L,\bar M)).
\end{align}

 %\end{equation}
 %\end{subequations}
%\begin{equation}\label{symmetry-lax}
%\begin{aligned}\frac{\d L}{\d
%s}&=\{-P_<(F(L,M)),L\}=\{P_\geqslant(F(L,M)),L\}-\{F(L,M),L\}\\&=
%\{P_\geqslant(F(L,M)),L\}+\frac{\partial F}{\partial M}(L,M)
%\end{aligned}
%\end{equation}
%
%
% %\eqref{symmetries-laxorlov}
%read as
% \begin{equation}\label{symmetries-laxorlov}
%\begin{aligned}
%\frac{\d L}{\d s}&=\{-P_<(F(L,M)),L\},
%&\frac{\d M}{\d s}&=\{-P_<(F(L,M)),M\},\\
%\frac{\d \bar L}{\d s}&=\{P_\geqslant(F(L,M)),\bar L\}, &\frac{\d
%\bar M}{\d s}&=\{P_\geqslant(F(L,M)),\bar M\}. \end{aligned}
% \end{equation}
%Let us concentrate in the analysis of
%\begin{equation}\label{symmetry-lax}
%\begin{aligned}\frac{\d L}{\d
%s}&=\{-P_<(F(L,M)),L\}=\{P_\geqslant(F(L,M)),L\}-\{F(L,M),L\}\\&=
%\{P_\geqslant(F(L,M)),L\}+\frac{\partial F}{\partial M}(L,M)
%\end{aligned}
%\end{equation}
%and also let us assume that
Hereon we shall assume that
\begin{equation}\label{F}
F(L,M)=\sum c_{ij}L^iM^j,\quad \bar F(\bar L,\bar M)=\sum \bar
c_{ij}\bar L^i\bar M^j.
\end{equation}

Let us keep  $t_n=0$ for $n>N$ and $\bar t_n=0$ for $n> \bar N$,
then recalling \eqref{MinL} we write
\begin{align}\label{MN}
M&=(N+1-r)t_NL^N+\dots+(2-r)t_1L+x+w_1L^{-1}+w_2 L^{-2}+\cdots,\\
\bar M&=-(r+\bar N-1)\bar t_{\bar N}\bar L^{-\bar N}-\dots-r\bar
t_1\bar L^{-1}+X+\bar w_1\bar L +\bar w_2\bar L^2+\cdots
\end{align}
Notice that if we want to keep $t_n=0$ for $n>N$ within the
transformation given by the symmetry, then \eqref{L} imply that
the function $F(L,M)$, when $M$ is expressed as in \eqref{MN}, has
no  terms proportional to $L^n$ for $n>N-r+1$. But as $F$ has the
form indicated in \eqref{F} we only need to impose this condition
over each of the products $L^iM^j$
\begin{align*}
L^iM^j&=L^i((N+1-r)t_NL^N+\dots+(2-r)t_1L+x+w_1L^{-1}+w_2
L^{-2}+\dots)^j\\
&=((N+1-r)t_N)^jL^{i+Nj}+\cdots\Rightarrow c_{ij}=0 \text{  if  }
i+Nj>N-r+1.
\end{align*}
Hence,
\begin{equation}\label{Fform}
F(L,M)=\sum_{n=1-r}^{N-r+1}\alpha_n\Big(\frac{M}{(N+1-r)L^N}\Big)L^n
\end{equation}
with $\alpha_n$ analytic functions.

The same reasoning may be applied in order to keep $\bar t_n=0$
for $n>\bar N$, and the corresponding condition is
\begin{equation}\label{barFform}
\bar F(\bar L,\bar M)=\sum_{n=1-r}^{\bar r-\bar N+1}\bar
\alpha_n\Big(\frac{\bar M}{(\bar N-1+r)\bar L^{-\bar N}}\Big)\bar
L^{-n}
\end{equation}
with $\bar \alpha_n$ analytic functions.

 \subsection{Symmetries of the potential $r$-dmKP equation}
In the following lines we shall find three symmetries of the
$r$-dmKP equation \eqref{mKP}. Let us suppose that $N=2$, and
$n=1-r,2-r$ and $3-r$, so that we have three different
contributions, or generators, to $F$, namely
\[
\alpha\big(\frac{M}{(3-r)L^2}\big)L^{1-r},\,\alpha\big(\frac{M}{(3-r)L^2}\big)L^{2-r}\text{
and } \alpha\big(\frac{M}{(3-r)L^2}\big)L^{3-r}.
\]
We first observe that
\[
\frac{M}{(3-r)L^2}=t_2+\frac{2-r}{3-r}t_1L^{-1}+\frac{1}{3-r}xL^{-2}-\frac{r}{3-r}\Psi_1
L^{-3}+\cdots
\]
If we denote
\[
\varepsilon:=\frac{2-r}{3-r}t_1L^{-1}+\frac{1}{3-r}xL^{-2}-\frac{r}{3-r}\Psi_1
L^{-3}+\cdots
\]
we have the following Taylor expansion
\begin{align*}
\alpha(t_2+\varepsilon)=&\alpha(t_2)+\dot\alpha(t_2)\varepsilon+\frac{1}{2}\ddot\alpha(t_2)\varepsilon^2
+\frac{1}{6}\dddot\alpha(t_2)\varepsilon^3+\cdots\\
=&\alpha(t_2)+\frac{2-r}{3-r}\dot\alpha(t_2)t_1L^{-1}+\Big(\frac{1}{3-r}\dot\alpha(t_2)x+\frac{(2-r)^2}{2(3-r)^2}
\ddot\alpha(t_2)t_1^2\Big)L^{-2}\\&+\Big(-\frac{r}{3-r}\dot\alpha(t_2)\Psi_1+\frac{2-r}{(3-r)^2}\ddot\alpha(t_2)t_1x
+\frac{(2-r)^3}{6(3-r)^3}\dddot\alpha(t_2)t_1^3\Big)L^{-3}+\cdots.
\end{align*}
We shall now analyze each of the three cases
\begin{enumerate}
\item Now, we have
\begin{align*}
F=\alpha(t_2+\varepsilon)L^{1-r}
=&\alpha(t_2)L^{1-r}+\frac{2-r}{3-r}\dot\alpha(t_2)t_1L^{-r}+\cdots.
\end{align*}
so that
\[
\frac{\d\psi_<}{\d
s}\cdot\psi_<^{-1}=-P_<F=-\alpha(t_2)P_<(L^{1-r})-\frac{2-r}{3-r}\dot\alpha(t_2)t_1L^{-r}+\cdots
\]
Recalling that
\[
\frac{\d\psi_<}{\d
s}\cdot\psi_<^{-1}=(\partial_s\Psi_1)p^{-r}+\big(\partial_s\Psi_2+
\frac{r}{2}(\Psi_{1,x}\partial_s\Psi_1-\Psi_1\partial_s\Psi_{1,x})\big)
p^{-r-1}+\cdots
%\mathcal{O}(p^{-r-2}).
\]
we deduce for $\Psi_1$ the following PDE
\[
\Psi_{1,s}=(1-r)\alpha(t_2)\Psi_{1,x}-\frac{2-r}{3-r}\dot\alpha(t_2)t_1
\]
whose solution is given by
\[
\Psi_1(s)=\frac{2-r}{(1-r)(3-r)}\frac{\dot\alpha(t_2)}{\alpha(t_2)}t_1x+f\Big(t_1,t_2,s+\frac{x}{(1-r)\alpha(t_2)}\Big)
\]
with $f$ and arbitrary function.

 For $s=0$ we obtain
\[
\Psi_1=\Psi_1(s)\Big|_{s=0}=\frac{2-r}{(1-r)(3-r)}\frac{\dot\alpha(t_2)}{\alpha(t_2)}t_1x+f\Big(t_1,t_2,\frac{x}{(1-r)\alpha(t_2)}\Big)
\]
from where we obtain
\begin{multline*}
\Psi_1(x+(1-r)s\alpha(t_2),t_1,t_2)=\frac{2-r}{(1-r)(3-r)}\frac{\dot\alpha(t_2)}{\alpha(t_2)}t_1(x+(1-r)s\alpha(t_2))\\+
f\Big(t_1,t_2,s+\frac{x}{(1-r)\alpha(t_2)}\Big).
\end{multline*}
Hence,
\[
\Psi_1(s)=-\frac{2-r}{3-r}s\dot\alpha(t_2)t_1+\Psi_1(x+(1-r)s\alpha(t_2),t_1,t_2).
\]
Therefore, we conclude that given any solution
$\Psi_1(x,t_1,t_2)$ of the potential $r$-dmKP equation \eqref{mKP}
and any analytic function $\alpha(t_2)$ then
\[
\tilde\Psi_1(x,t_1,t_2):=-\frac{2-r}{3-r}\dot\alpha(t_2)t_1+\Psi_1(x+(1-r)\alpha(t_2),t_1,t_2).
\]
is a new solution of the equation. For $u=-\Psi_{1,x}$ the
symmetry transformation for a solution of the $r$-dmKP equation
\eqref{r-dmKP} is given by
\[
\tilde u(x,t_1,t_2)=u(x+(1-r)\alpha(t_2),t_1,t_2),
\]
which for $r=1$ is the identity transformation.

\item In this case
\begin{align*}
F=&\alpha(t_2+\varepsilon)L^{2-r}\\
=&\alpha(t_2)L^{2-r}+\frac{2-r}{3-r}\dot\alpha(t_2)t_1L^{1-r}+\Big(\frac{1}{3-r}\dot\alpha(t_2)x+\frac{(2-r)^2}{2(3-r)^2}
\ddot\alpha(t_2)t_1^2\Big)L^{-r}+\cdots
\end{align*}
so that
\begin{multline*}
\frac{\d\psi_<}{\d
s}\cdot\psi_<^{-1}=\alpha(t_2)\frac{\d\psi_<}{\d
t_1}\cdot\psi_<^{-1}-\frac{2-r}{3-r}\dot\alpha(t_2)t_1P_<L^{1-r}\\-\Big(\frac{1}{3-r}\dot\alpha(t_2)x+\frac{(2-r)^2}{2(3-r)^2}
\ddot\alpha(t_2)t_1^2\Big)L^{-r}+\cdots.
\end{multline*}

For $\Psi_1$ we find the following PDE
\[
\Psi_{1,s}=\alpha(t_2)\Psi_{1,t_1}+\frac{(1-r)(2-r)}{3-r}\dot\alpha(t_2)t_1\Psi_{1,x}-\frac{1}{3-r}\dot\alpha(t_2)x-\frac{(2-r)^2}{2(3-r)^2}
\ddot\alpha(t_2)t_1^2
\]
whose solution is given by
\begin{equation}\label{sol2}
\Psi_1(s)=g(x,t_1,t_2)+f\Big(t_2,t_1^2-\frac{2(3-r)}{(1-r)(2-r)}\frac{\alpha}{\dot\alpha}x,s+\frac{t_1}{\alpha}\Big)
\end{equation}
being $f$ an arbitrary function and \[
g:=\frac{1}{3-r}\frac{\dot\alpha}{\alpha}xt_1+\frac{(2-r)}{6(3-r)^2}
\Big((2-r)\frac{\ddot\alpha}{\alpha}-
2(1-r)\frac{\dot\alpha^2}{\alpha^2}\Big)t_1^3.
\]
 Setting $s=0$ in \eqref{sol2} we arrive to
\begin{equation}\label{sol20}
\Psi_1=\Psi_1\big|_{s=0}=g(x,t_1,t_2)
+f\Big(t_2,t_1^2-\frac{2(3-r)}{(1-r)(2-r)}\frac{\alpha}{\dot\alpha}x,\frac{t_1}{\alpha}\Big).
\end{equation}

If in \eqref{sol20} we  replace the independent variables $x$ and
$t_1$ by
\begin{equation*}
\begin{aligned}
x(s)&= x+s\Big(\frac{(1-r)(2-r)}{2(3-r)}\dot\alpha(s\alpha+2t_1)\Big),\\
\tilde t_1(s)&= t_1+s\alpha,
\end{aligned}
\end{equation*}
we deduce
\[
f\big(t_2,t_1^2-\frac{2(3-r)}{(1-r)(2-r)}\frac{\alpha}{\dot\alpha}x,s+\frac{t_1}{\alpha}\big)=\Psi_1(x(s),
t_1(s),t_2)-g(x(s),t_1(s),t_2).
\]

Hence, from \eqref{sol2} we infer that
\[
\Psi_1(s)=g(x,t_1,t_2)-g(x(s),t_1(s),t_2)+\Psi_1(x(s),t_1(s),
t_2),
\]
and therefore
\begin{multline}\label{symmetry-s-2}
\Psi_1(s)=\Psi_1\Big(x+s\frac{(1-r)(2-r)}{2(3-r)}\dot\alpha(s\alpha+2t_1),t_1+s\alpha,t_2\Big)\\
-s\frac{1}{3-r}\dot\alpha x- s\frac{(2-r)^2}{2(3-r)^2}\ddot\alpha
t_1^2-s^2\frac{2-r}{6(3-r)^2}((1-r)\dot\alpha^2+(2-r)\alpha\ddot\alpha)(3t_1+s\alpha).
\end{multline}

Hence, given any solution $\Psi_1(x,t_1,t_2)$ of the potential
$r$-dmKP equation \eqref{mKP} and any analytic function
$\alpha(t_2)$ then
\begin{equation}\label{symmetry-2}
\begin{aligned}
\tilde\Psi_1=&\Psi_1\Big(x+\frac{(1-r)(2-r)}{2(3-r)}\dot\alpha(\alpha+2t_1),t_1+\alpha,t_2\Big)\\
&-\frac{1}{3-r}\dot\alpha x- \frac{(2-r)^2}{2(3-r)^2}\ddot\alpha
t_1^2-\frac{2-r}{6(3-r)^2}((1-r)\dot\alpha^2+(2-r)\alpha\ddot\alpha)(3t_1+\alpha)
\end{aligned}
\end{equation}
is a new solution of \eqref{mKP}.

For $u=-\Psi_{1,x}$, thus $u$ solves the $r$-dmKP equation
\eqref{r-dmKP}, the corresponding symmetry transformation is given
by
\[
\tilde
u(x,t_1,t_2)=u\Big(x+\frac{(1-r)(2-r)}{2(3-r)}\dot\alpha(t_2)(\alpha(t_2)+2t_1),t_1+\alpha(t_2),t_2\Big)
+\frac{1}{3-r}\dot\alpha(t_2),
\]
which for $r=1$ simplifies to
\[
\tilde
u(x,t_1,t_2)=u(x,t_1+\alpha(t_2),t_2)+\frac{1}{2}\dot\alpha(t_2).
\]

\item Finally, we tackle the most involve case
\begin{align*}
F=&\alpha(t_2+\varepsilon)L^{3-r}\\
=&\alpha(t_2)L^{3-r}+\frac{2-r}{3-r}\dot\alpha(t_2)t_1L^{2-r}+\Big(\frac{1}{3-r}\dot\alpha(t_2)x+\frac{(2-r)^2}{2(3-r)^2}
\ddot\alpha(t_2)t_1^2\Big)L^{1-r}\\
&+\Big(-\frac{r}{3-r}\dot\alpha(t_2)\Psi_1+\frac{2-r}{(3-r)^2}\ddot\alpha(t_2)t_1x
+\frac{(2-r)^3}{6(3-r)^3}\dddot\alpha(t_2)t_1^3\Big)L^{-r}+\dots
\end{align*}
so that
\begin{equation}\label{psi-symmetry-30}
\begin{aligned}
\frac{\d\psi_<}{\d
s}\cdot\psi_<^{-1}=&\alpha(t_2)\frac{\d\psi_<}{\d
t_2}\cdot\psi_<^{-1}+\frac{2-r}{3-r}\dot\alpha(t_2)t_1\frac{\d\psi_<}{\d
t_1}\cdot\psi_<^{-1}\\&-\Big(\frac{1}{3-r}\dot\alpha(t_2)x+\frac{(2-r)^2}{2(3-r)^2}
\ddot\alpha(t_2)t_1^2\Big)P_<L^{1-r}\\&-\Big(-\frac{r}{3-r}\dot\alpha(t_2)\Psi_1+\frac{2-r}{(3-r)^2}\ddot\alpha(t_2)t_1x
+\frac{(2-r)^3}{6(3-r)^3}\dddot\alpha(t_2)t_1^3\Big)L^{-r}+\cdots.
\end{aligned}
\end{equation}
From \eqref{psi-symmetry-30} we deduce that $\Psi_1(s)$ solves the
following PDE
\begin{equation}\label{psi-symmetry-3}
\begin{aligned}
\Psi_{1,s}=&\alpha\Psi_{1,t_2}+\frac{2-r}{3-r}\dot\alpha
t_1\Psi_{1,t_1} +(1-r)\Big(\frac{1}{3-r}\dot\alpha
x+\frac{(2-r)^2}{2(3-r)^2} \ddot\alpha
t_1^2\Big)\Psi_{1,x}\\&+\frac{r}{3-r}\dot\alpha\Psi_1-\frac{2-r}{(3-r)^2}\ddot\alpha
t_1x -\frac{(2-r)^3}{6(3-r)^3}\dddot\alpha t_1^3.
\end{aligned}
\end{equation}
The general solution of \eqref{psi-symmetry-3} is
\begin{multline}\label{psi-symmetry-3-sol}
\Psi_1(s)=\alpha^{-\frac{r}{3-r}}f\Big(t_1\alpha^{-\frac{2-r}{3-r}},
x\alpha^{-\frac{1-r}{3-r}}-\frac{(1-r)(2-r)^2}{2(3-r)^2}t_1^2\alpha^{-2\frac{2-r}{3-r}}\dot\alpha,s+\int^{t_2}\frac{\d
t}{\alpha(t)}\Big)\\+g(x,t_1,t_2)
\end{multline}
where $f$ is an arbitrary function and $g$ is given by
\begin{gather*}
g(x,t_1,t_2):= \frac{2-r}{(3-r)^2}\frac{\dot\alpha}{\alpha}
xt_1+\frac{(2-r)^3}{6(3-r)^4}\
\bigg((3-r)\frac{\ddot\alpha}{\alpha}
-(3-2r)\frac{\dot\alpha^2}{\alpha^2}\bigg)t_1^3.
\end{gather*}
%or
%\begin{multline*}
%g(x,t_1,t_2):= \frac{2-r}{(3-r)^2}\Big(\frac{\d\log\alpha}{\d t_2}
%xt_1+\frac{(2-r)^2}{6(3-r)}\ \bigg(\frac{\d^2\log\alpha}{\d
%t_2^2}-\frac{3-2r}{3-r}\big(\frac{\d\log\alpha}{\d
%t_2}\big)^2\bigg)t_1^3\Big).
%%\frac{2-r}{(3-r)^2}\frac{\dot\alpha}{\alpha}
%%xt_1+\frac{(2-r)^3}{6(3-r)^4}\
%%\bigg((3-r)\frac{\ddot\alpha}{\alpha}
%%-(3-2r)\frac{\dot\alpha^2}{\alpha^2}\bigg)t_1^3.
%\end{multline*}

We define
\[
c(t)=\int^{t}\frac{\d t}{\alpha(t)}
\]
and define $T$ by the relation
\[
c(T)=s+c(t_2),
 \]
or
\begin{equation}\label{T}
\int_{t_2}^T\frac{\d t}{\alpha(t)}=s.
\end{equation}
Observe that from \eqref{T} we derive
\begin{equation}\label{DT}
\frac{\d T}{\d
t_2}\frac{1}{\alpha(T)}-\frac{1}{\alpha(t_2)}=0\Rightarrow\frac{\d
T}{\d t_2}=\frac{\alpha(T(t_2))}{\alpha(t_2)}
\end{equation}
and hence
\begin{equation}\label{DDT}
\frac{\d^2 T}{\d t_2^2}=\frac{\d T}{\d
t_2}\frac{\dot\alpha(T)-\dot\alpha(t_2)}{\alpha(t_2)}.
\end{equation}

 Now, we can write \eqref{psi-symmetry-3-sol} as
\begin{gather*}
\Psi_1(s)=\alpha^{-\frac{r}{3-r}}f\Big(t_1\alpha^{-\frac{2-r}{3-r}},
x\alpha^{-\frac{1-r}{3-r}}-\frac{(1-r)(2-r)^2}{2(3-r)^2}t_1^2\alpha^{-2\frac{2-r}{3-r}}
\dot\alpha,c(T)\Big)+g(x,t_1,t_2),
\end{gather*}
which setting $s=0$ reads
\begin{gather*}
\Psi_1=\alpha^{-\frac{r}{3-r}}f\Big(t_1\alpha^{-\frac{2-r}{3-r}},
x\alpha^{-\frac{1-r}{3-r}}-\frac{(1-r)(2-r)^2}{2(3-r)^2}t_1^2\alpha^{-2\frac{2-r}{3-r}}
\dot\alpha,c(t_2)\Big)+g(x,t_1,t_2).
\end{gather*}

Then, we introduce the following curve
\begin{align*} t_2(s):=&T(t_2),\\
t_1(s):=&\Big(\frac{\alpha(t_2)}{\alpha(T(t_2))}\Big)^{-\frac{2-r}{3-r}}t_1,\\
x(s):=&\Big(\frac{\alpha(t_2)}{\alpha(T(t_2))}\Big)^{-\frac{1-r}{3-r}}x
\\&+\frac{(1-r)(2-r)^2}{2(3-r)^2}\alpha(T(t_2))^{\frac{1-r}{3-r}}\alpha(t_2)^{-2\frac{2-r}{3-r}}\big(\dot\alpha(T(t_2))
-\dot\alpha(t_2)\big)t_1^2
\end{align*}
which using  \eqref{DT} and \eqref{DDT} can expressed as
\begin{equation}\label{curve}\begin{aligned} t_2(s):=&T(t_2),\\
t_1(s):=&(\dot T(t_2))^{\frac{2-r}{3-r}}t_1,\\
x(s):=&(\dot
T(t_2))^{\frac{1-r}{3-r}}\Big(x+\frac{(1-r)(2-r)^2}{2(3-r)^2}\frac{\ddot
T(t_2)}{\dot T(t_2)}t_1^2\Big),
\end{aligned}
\end{equation}
With the curve parameterized as in \eqref{curve} we see that
\begin{multline*}
f\Big(t_1\alpha^{-\frac{2-r}{3-r}},
x\alpha^{-\frac{1-r}{3-r}}-\frac{(1-r)(2-r)^2}{2(3-r)^2}t_1^2\dot\alpha,s+\int^{t_2}\frac{\d
t}{\alpha(t)}\Big)\\=\alpha(T)^{\frac{r}{3-r}}(\Psi_{1}(x(s),t_1(s),t_2(s))
-g(x(s),t_1(s),t_2(s))),
\end{multline*}
and therefore
\begin{equation}\label{symmetry-pre}
\Psi_1(s)=g(x,t_1,t_2)-\dot
T^{\frac{r}{3-r}}g(x(s),t_1(s),t_2(s))+\dot
T^{\frac{r}{3-r}}\Psi_1(x(s),t_1(s),t_2(s)).
\end{equation}
Let us evaluate
\[
g(x,t_1,t_2)-\dot
T^{\frac{r}{3-r}}g(x(s),t_1(s),t_2(s))=A(t_2)xt_1+B(t_2)t_1^3
\]
where
\begin{align}
\label{A0}A:=&\frac{2-r}{(3-r)^2}\Big(\frac{\dot\alpha}{\alpha}-\dot
T\frac{\dot\alpha(T)}{\alpha(T)}\Big)
\\
\label{B0}B:=&\frac{(2-r)^3}{6(3-r)^4}\Big((3-r)\Big(\frac{\ddot\alpha}{\alpha}-\dot
T\frac{\ddot
\alpha(T)}{\alpha(T)}\Big)-(3-2r)\Big(\frac{\dot\alpha^2}{\alpha^2}-\dot
T^2\frac{\dot
\alpha(T)^2}{\alpha(T)^2}\Big)\Big)\\&-\frac{(2-r)^3}{2(3-r)^4}(1-r)\ddot
T\frac{\dot\alpha(T)}{\alpha(T)}.\notag
\end{align}
From \eqref{DT}, \eqref{DDT} and \eqref{A0} we derive
\begin{equation}\label{A}
A=-\frac{2-r}{(3-r)^2}\frac{\ddot T}{\dot T}.
\end{equation}
A similar expression may be derived,  using \eqref{DT},
\eqref{DDT} and its consequences, for $B$ solely in terms of $T$
and its derivatives. However a faster way is to reckon
$Axt_1+Bt_1^3$ as a solution of \eqref{mKP} ---just by applying
the symmetry to $\Psi_1=0$---. In doing so we find that
\[
B=-\frac{(2-r)^2}{6(3-r)}(\dot A+r(3-r)A^2)
\]
and hence, by taking into account \eqref{A} we deduce the
following expression for $B$ in terms solely of $T$ and its
derivatives
\begin{equation}\label{B}
B=-\frac{(2-r)^3}{6(3-r)^3}\Big(\frac{\dddot T}{\dot
T}-\frac{3}{3-r}\,\frac{\ddot T^2}{\dot T^2}\Big).
\end{equation}
Collecting \eqref{symmetry-pre} together with \eqref{curve},
\eqref{A} and \eqref{B} we deduce the following: If $
\Psi(x,t_1,t_2)$ is a solution of the potential $r$-dmKP equation
\eqref{mKP} and $T(t_2)$ is an arbitrary  function of $t_2$ then
\begin{equation}\label{symmetry3}
\begin{aligned}
\tilde \Psi_1(x,t_1,t_2)=&-\frac{2-r}{(3-r)^2}\frac{\ddot T}{\dot
T}xt_1-\frac{(2-r)^3}{6(3-r)^3}\Big(\frac{\dddot T}{\dot
T}-\frac{3}{3-r}\,\frac{\ddot T^2}{\dot T^2}\Big)t_1^3\\&+ \dot
T^{\frac{r}{3-r}}\Psi_1\Big(\dot
T^{\frac{1-r}{3-r}}\Big(x+\frac{(1-r)(2-r)^2}{2(3-r)^2}\frac{\ddot
T}{\dot T}t_1^2\Big), \dot T^{\frac{2-r}{3-r}}t_1,T\Big)
\end{aligned}
\end{equation}
is a new solution of \eqref{mKP}. As previously for
$u=-\Psi_{1,x}$ we have: given a solution $u$ of the $r$-dmKP
equation \eqref{r-dmKP} and an arbitary function $T(t_2)$ then
$\tilde u$ defined by
\begin{equation}\label{symmetry3-mKP}
\tilde u =\frac{2-r}{(3-r)^2}\frac{\ddot T}{\dot T}x +\dot
T^{\frac{1}{3-r}}u\Big(\dot
T^{\frac{1-r}{3-r}}\Big(x+\frac{(1-r)(2-r)^2}{2(3-r)^2}\frac{\ddot
T}{\dot T}t_1^2\Big), \dot T^{\frac{2-r}{3-r}}t_1,T\Big)
\end{equation}
is a new solution of \eqref{r-dmKP}.
\end{enumerate}
We collect these results regarding the potential $r$-dmKP equation
in the following

\begin{pro}\label{symmeries-rdmkp}
Given a solution $\Psi_1$ of the potential $r$-dmKP equation
\begin{gather*}
\Psi_{1,xt_2}=\frac{3-r}{(2-r)^2}\Psi_{1,t_1t_1}-\frac{(3-r)(1-r)}{2-r}\Psi_{1,xx}\Psi_{1,t_1}-
\frac{(3-r)r}{2-r}\Psi_{1,x}\Psi_{1,xt_1}-\frac{(3-r)(1-r)}{2}\Psi_{1,x}^2\Psi_{1,xx}
\end{gather*}
and arbitrary functions $\alpha(t_2)$, $T(t_2)$ the following
functions are new solutions of the $r$-dmKP equation:
\begin{align*}
\tilde\Psi_1=&-\frac{2-r}{3-r}\dot\alpha(t_2)t_1+\Psi_1(x+(1-r)\alpha(t_2),t_1,t_2),\\
\tilde\Psi_1= &-\frac{1}{3-r}\dot\alpha(t_2) x-
\frac{(2-r)^2}{2(3-r)^2}\ddot\alpha(t_2)
t_1^2-\frac{2-r}{6(3-r)^2}((1-r)\dot\alpha(t_2)^2+(2-r)\alpha(t_2)
\ddot\alpha(t_2))(3t_1+\alpha(t_2))\\
&+\Psi_1\Big(x+\frac{(1-r)(2-r)}{2(3-r)}\dot\alpha(t_2)(\alpha(t_2)+2t_1),t_1
+\alpha(t_2),t_2\Big)\\[8pt]
\tilde \Psi_1=&-\frac{2-r}{(3-r)^2}\frac{\ddot T(t_2)}{\dot
T(t_2)}xt_1-\frac{(2-r)^3}{6(3-r)^3}\Big(\frac{\dddot T(t_2)}{\dot
T(t_2)}-\frac{3}{3-r}\,\frac{\ddot T(t_2)^2}{\dot
T(t_2)^2}\Big)t_1^3\\&+ \dot T(t_2)^{\frac{r}{3-r}}\Psi_1\Big(\dot
T(t_2)^{\frac{1-r}{3-r}}\Big(x+\frac{(1-r)(2-r)^2}{2(3-r)^2}\frac{\ddot
T(t_2)}{\dot T(t_2)}t_1^2\Big), \dot
T(t_2)^{\frac{2-r}{3-r}}t_1,T(t_2)\Big).
\end{align*}

\end{pro}

A similar proposition  for the $r$-dmKP equation \eqref{r-dmKP}
follows

\begin{pro}\label{pro-symmetries-dmKP}
Given a solution $u$ of the  $r$-dmKP equation
\begin{multline*}
u_{t_2}=\frac{3-r}{(2-r)^2}(\partial_x^{-1}u)_{t_1t_1}+
\frac{(3-r)(1-r)}{2-r}u_x(\partial_x^{-1}u)_{t_1}+\frac{r(3-r)}{2-r}uu_{t_1}-
\frac{(3-r)(1-r)}{2}u^2u_x.
\end{multline*}
and arbitrary functions $\alpha(t_2)$, $T(t_2)$ the following
functions are new solutions of the $r$-dmKP equation:
\begin{align*}
\tilde u=& u(x+(1-r)\alpha(t_2),t_1,t_2),\\
\tilde u=
&\frac{1}{3-r}\dot\alpha(t_2)+u\Big(x+\frac{(1-r)(2-r)}{2(3-r)}\dot\alpha(t_2)(\alpha(t_2)+2t_1),t_1
+\alpha(t_2),t_2\Big)\\[8pt]
\tilde u=&\frac{2-r}{(3-r)^2}\frac{\ddot T(t_2)}{\dot T(t_2)}t_1+
\dot T(t_2)^{\frac{1}{3-r}}u\Big(\dot
T(t_2)^{\frac{1-r}{3-r}}\Big(x+\frac{(1-r)(2-r)^2}{2(3-r)^2}\frac{\ddot
T(t_2)}{\dot T(t_2)}t_1^2\Big), \dot
T(t_2)^{\frac{2-r}{3-r}}t_1,T(t_2)\Big).
\end{align*}

\end{pro}

\subsection{Symmetries of the potential $r$-dDym equation}
 We shall find three symmetries for the $r$-dDym equation \eqref{PDEbarX}.
 From \eqref{symmetries>} we deduce that
 \[
 \frac{\d \bar\psi_{1-r}}{\d
 s}\cdot\bar\psi_{1-r}^{-1}+\Ad_{\bar\psi_{1-r}}P_{1-r}F(L,M)=0
 \]
 where
 \[
 \frac{\d \bar\psi_{1-r}}{\d
 s}\cdot\bar\psi_{1-r}^{-1}=\left\{
 \begin{aligned}&\frac{1}{1-r}\frac{\bar X_s}{\bar X_x}p^{1-r}, & r&\neq 1,\\
&-\xi_sp^{1-r}, & r&=1.
 \end{aligned}\right.
 \]
As for the previous case we pay particular attention to  the case
$N=2$, with $n=1-r,2-r$ and $3-r$, so that $F$ can be taken as
\[
\alpha\big(\frac{M}{(3-r)L^2}\big)L^{1-r},\,\alpha\big(\frac{M}{(3-r)L^2}\big)L^{2-r}\text{
and } \alpha\big(\frac{M}{(3-r)L^2}\big)L^{3-r},
\]
and
\begin{align*}
\alpha\big(\frac{M}{(3-r)L^2}\big)
=&\alpha(t_2)+\frac{2-r}{3-r}\dot\alpha(t_2)t_1L^{-1}+\Big(\frac{1}{3-r}\dot\alpha(t_2)x+\frac{(2-r)^2}{2(3-r)^2}
\ddot\alpha(t_2)t_1^2\Big)L^{-2}\\&+\Big(-\frac{r}{3-r}\dot\alpha(t_2)\Psi_1+\frac{2-r}{(3-r)^2}\ddot\alpha(t_2)t_1x
+\frac{(2-r)^3}{6(3-r)^3}\dddot\alpha(t_2)t_1^3\Big)L^{-3}+\cdots.
\end{align*}
\begin{enumerate}
\item Let us take $n=1-r$ and
\[
P_{1-r}F(L,M)=\alpha(t_2) p^{1-r},
\]
so that
\[
\Ad_{\bar\psi_{1-r}}P_{1-r}F(L,M)=\left\{
\begin{aligned}
&\frac{\alpha(t_2)}{\bar X_x}p^{1-r},& r&\neq 1,\\
&\alpha(t_2)p^{1-r},& r&=1.
\end{aligned}\right.
\]
Then, for $r\neq 1$ we have
\[
\bar X_s=-(1-r)\alpha(t_2)
\]
and
\[
\bar X(s)=\bar X- s(1-r)\alpha(t_2).
\]
Hence, given a solution $X$ to the potential $r$-dDym equation
\eqref{PDEbarX} and an arbitrary function $\alpha(t_2)$ the
function
\[
\tilde{\bar{X}}=\bar X-(1-r)\alpha(t_2),
\]
is a new solution of the potential $r$-dDym equation.

When $r=1$ we have
\[
\xi_s=\alpha\Rightarrow\xi(s)=\xi+\alpha(t_2)s.
\]

The corresponding symmetry of  \eqref{r-dDym} is the identity
transformation.

\item In this case   we set $n=2-r$ so that
\[
P_{1-r}F(L,M)=\alpha(t_2)
P_{1-r}L^{2-r}+\frac{2-r}{3-r}\dot\alpha(t_2)t_1p^{1-r}.
\]
Recall that
\[
\partial_n\bar\psi_{1-r}\cdot\bar\psi_{1-r}^{-1}+\Ad_{\bar\psi_{1-r}}P_{1-r}L^{n+1-r}=0
\]
and therefore
\begin{gather}\label{pepe}
\bar X_s=\alpha(t_2)\bar
X_{t_1}-\frac{(1-r)(2-r)}{3-r}\dot\alpha(t_2)t_1.
\end{gather}
The general solution of \eqref{pepe} is
\begin{equation}\label{solbarX}
\bar X(s)=\frac{(1-r)(2-r)}{2(3-r)}\frac{\dot\alpha}{\alpha}t_1^2
+f\Big(x,t_2,s+\frac{t_1}{ \alpha}\Big) \end{equation}
 where $f$
is an arbitrary function. Setting $s=0$ in \eqref{solbarX} we
obtain
\[
\bar X=\frac{(1-r)(2-r)}{2(3-r)}\frac{\dot\alpha}{\alpha}t_1^2
+f\Big(x,t_2,\frac{t_1}{ \alpha}\Big).
\]
If
\[
t_1(s)=\alpha s+t_1
\]
then
\[
f\Big(x,t_2,s+\frac{t_1}{ \alpha}\Big)=\bar
X(x,t_1(s),t_2)-\frac{(1-r)(2-r)}{2(3-r)}\frac{\dot\alpha}{\alpha}t_1(s)^2
\]
so that
\[
\bar X(s)=-\frac{(1-r)(2-r)}{2(3-r)}\dot\alpha(\alpha
s+2t_1)s+\bar X(x, t_1+s \alpha,t_2).
\]
Hence, given a solution $\bar X$ of the potential $r$-dDym
equation \eqref{PDEbarX} and an arbitrary function $\alpha(t_2)$
 the function $\tilde{\bar X}$ given by
\[
\tilde{\bar
X}=-\frac{(1-r)(2-r)}{2(3-r)}\dot\alpha(\alpha+2t_1)+\bar X(x,
t_1+ \alpha,t_2)
\]
is a new solution.

When $r=1$ we obtain
\[
\xi_s=\alpha \xi_{t_1}+\frac{\dot\alpha}{2}
\]
whose solution is
\[
\xi(s)=\frac{1}{4}\frac{\dot\alpha}{\alpha}t_1^2+f(x,t_1+s\alpha,t_2)
\]
and the transformation is given by
\[
\tilde\xi=\frac{1}{4}\dot\alpha(\alpha+2
t_1)+\xi(x,t_1+\alpha,t_2).
\]

The corresponding symmetry for the $r$-dDym equation
\eqref{r-dDym} is
\[
\tilde v=v(x,t_1+\alpha(t_2),t_2)
\]

 \item Now we set $n=3-r$ so that
\begin{align*}
P_{1-r}F(L,M)=&\alpha(t_2)
P_{1-r}L^{3-r}+\frac{2-r}{3-r}\dot\alpha(t_2)t_1P_{1-r}L^{2-r}\\&+
\Big(\frac{1}{3-r}\dot\alpha(t_2)x+\frac{(2-r)^2}{2(3-r)^2}
\ddot\alpha(t_2)t_1^2\Big)p^{1-r}
\end{align*}
and $\bar X(s)$ solves
\begin{gather}\label{pepe2}
\bar X_s=\alpha \bar X_{t_2}+\frac{2-r}{3-r}\dot\alpha t_1\bar
X_{t_1}-\frac{1-r}{3-r}\dot\alpha X-\frac{(1-r)(2-r)^2}{2(3-r)^2}
\ddot\alpha t_1^2.
\end{gather}
The general solution \eqref{pepe2} is
\[
\bar
X(s)=g(t_1,t_2)+\alpha^{-\frac{1-r}{3-r}}f\Big(x,t_1\alpha^{-\frac{2-r}{3-r}},
s+\int^{t_2}\frac{\d t}{\alpha(t)}\Big)
\]
where $f$ is an arbitrary function and
\[
g(t_1,t_2):= \frac{(1-r)(2-r)^2}{2(3-r)^2}t_1^2
\frac{\dot\alpha}{\alpha}.
\]
We now proceed as before; i. e., we
define $c$ and $T$ as follows
\[
c(t_2)=\int^{t_2}\frac{\d t}{\alpha(t)},\quad c(T)=s+c(t_2),\quad
\int_{t_2}^T\frac{\d t}{\alpha(t)}=s.
\]
For $s=0$ we have
\[
\bar
X=g(t_1,t_2)+\alpha^{-\frac{1-r}{3-r}}F(x,t_1\alpha^{-\frac{2-r}{3-r}},c(t_2))
\]
so that defining the curve
\begin{align*}
t_1(s)&=\frac{\alpha(T)^{\frac{2-r}{3-r}}}{\alpha(t_2)^{\frac{2-r}{3-r}}}t_1=
\dot T^{\frac{2-r}{3-r}}t_1,\\
t_2(s)&=T
\end{align*}
we have
\[
f(x,t_1\alpha(t_2)^{-\frac{2-r}{3-r}},c(T))=\alpha(T)^{-\frac{1-r}{3-r}}\big(X(x,t_1(s),
t_2(s))-g(t_1(s),t_2(s))\big)
\]
and hence
\[
\tilde X(s)=g(t_1,t_2)-\dot
T^{-\frac{1-r}{3-r}}g(t_1(s),t_2(s))+\dot T^{-\frac{1-r}{3-r}}\bar
X(x,t_1(s),t_2(s))
\]

Let us analyze
\begin{align*}
g(t_1,t_2)-g(t_1(s),t_2(s))=&\frac{(1-r)(2-r)^2}{2(3-r)^2}t_1^2
\Big(\frac{\dot\alpha(t_2)}{\alpha(t_2)}-\dot
T\frac{\dot\alpha(T)}{\alpha(T)}\Big)\\
=&-\frac{(1-r)(2-r)^2}{2(3-r)^2}t_1^2 \frac{\ddot T}{\dot T},
\end{align*}
expression that allows us to write
\begin{equation}\label{symmetry-dym3}
\bar X(s)=-\frac{(1-r)(2-r)^2}{2(3-r)^2}t_1^2 \frac{\ddot T}{\dot
T}+ \dot T^{-\frac{1-r}{3-r}}\bar X(x,\dot
T^{\frac{2-r}{3-r}}t_1,T).
\end{equation}

Thus, given a solution $\bar X$ of the potential $r$-dDym equation
\eqref{PDEbarX} and arbitrary function $T(t_2)$ a new solution
$\tilde{\bar X}=\bar X(s)$ is given by \eqref{symmetry-dym3}. The
corresponding symmetry of the $r$-dDym equation \eqref{r-dDym} is
\[
\tilde v=\dot T^{\frac{1}{3-r}}v(x,\dot
T(t_2)^{\frac{2-r}{3-r}}t_1,T(t_2))
\]

When $r=1$ we find for $\xi$ the following PDE
\[
\xi_s=\alpha\xi_{t_2}+\frac{\dot\alpha}{2}t_1\xi_{t_1}+\frac{\dot\alpha}{2}x+
\frac{\ddot\alpha}{8}t_1^2
\]
whose solution is
\[
\xi(s)=-\frac{x}{2}\log\alpha-\frac{1}{8}\frac{\dot\alpha}{\alpha}t_1^2+F
\Big(x,\frac{1}{\sqrt{\alpha}}t_1+s+\int\frac{\d
t}{\alpha(t)}\Big).
\]
Finally, from this formula we derive that
\[
\tilde\xi=\frac{x}{2}\log(\dot T(t_2))+\frac{t_1^2}{8}\frac{\ddot
T(t_2)}{\dot T(t_2)}+\xi\big(x,\sqrt{\dot T(t_2)}t_1,T(t_2)\big)
\]
is a new solution.

\end{enumerate}

We gather all these results regarding the potential $r$-dDym
equation in the following
\begin{pro}\label{symmetries-prddym}
Given a solution $\bar X$ of the potential $r$-dDym equation
($r\neq 1$)
\[
\bar X_{xt_2}=\frac{3-r}{2-r}\Big(\frac{1}{2-r}\bar X_{t_1t_1}\bar
X_x-\frac{1}{1-r}\bar X_{xt_1}\bar X_{t_1}\Big),
\]
and arbitrary functions $\alpha(t_2)$, $T(t_2)$ the following
functions are new solutions of the potential $r$-dDym equation
($r\neq 1$):
\begin{align*}
\tilde{\bar X}=& \bar X(x,t_1,t_2)-(1-r)\alpha(t_2),\\
\tilde{\bar
X}=&-\frac{(1-r)(2-r)}{2(3-r)}\dot\alpha(t_2)(\alpha(t_2)+2t_1)+\bar
X(x,
t_1+ \alpha(t_2),t_2),\\
\tilde{\bar X}=& -\frac{(1-r)(2-r)^2}{2(3-r)^2}t_1^2 \frac{\ddot
T(t_2)}{\dot T(t_2)}+ \dot T(t_2)^{-\frac{1-r}{3-r}}\bar X(x,\dot
T(t_2)^{\frac{2-r}{3-r}}t_1,T(t_2)).
\end{align*}

Given a solution $\xi$ of the  potential $r=1$ dDym equation
\[
\xi_{t_2x}-2\xi_{t_1t_1}-2\xi_{t_1}\xi_{t_1x}=0
\]
new solutions $\tilde\xi$ are given by
\begin{align*}
\tilde\xi=&\alpha(t_2)+ \xi(x,t_1,t_2),\\
\tilde\xi=&\frac{1}{4}\dot\alpha(t_2)(\alpha(t_2)+2t_1)+\xi(x,
t_1+ \alpha(t_2),t_2),\\
\tilde\xi=& \frac{x}{2}\log(\dot
T(t_2))+\frac{t_1^2}{8}\frac{\ddot T(t_2)}{\dot
T(t_2)}+\xi\big(x,\sqrt{\dot T(t_2)}t_1,T(t_2)\big).
\end{align*}
\end{pro}

We also resume the results for the $r$-dDym equation
\eqref{r-dDym}

\begin{pro}
Given a solution $v$ of the $r$-dDym equation
\[
v_{t_2}=\frac{3-r}{(2-r)^2}v^{r-1}\Big(v^{2-r}\partial_x^{-1}(v^{r-2}v_{t_1})\Big)_{t_1},
\]
and arbitrary functions $\alpha(t_2)$, $T(t_2)$ the following
functions are new solutions of the $r$-dDym equation:
\begin{align*}
\tilde v=& v(x,t_1+\alpha(t_2),t_2),\\
\tilde v=& \dot T(t_2)^{\frac{1}{3-r}}v(x,\dot
T(t_2)^{\frac{2-r}{3-r}}t_1,T(t_2)).
\end{align*}
\end{pro}

\subsection{Symmetries of the $r$-dToda equation}

The $r$-dToda equation \eqref{toda} mixes the independent
variables $x$, $t_1$ and $\bar t_1$. Let us analyze the symmetries
associated with the $t_1$-flow; i. e., study the action of
additional symmetries generated by $F(L,M)$. Suppose that $N=1$,
then we have the cases $n=1-r$ and $2-r$, so that we have two
different generators, namely
\[
\alpha\big(\frac{M}{(2-r)L}\big)L^{1-r}\text{ and
}\alpha\big(\frac{M}{(2-r)L}\big)L^{2-r}.
\]
We first observe that
\[
\frac{M}{(2-r)L^2}=t_1+\frac{1}{2-r}xL^{-1}-\frac{r}{2-r}\Psi_1
L^{-2}+\cdots
\]
If we denote
\[
\varepsilon:=\frac{1}{2-r}xL^{-1}-\frac{r}{2-r}\Psi_1
L^{-2}+\cdots
\]
we have the following Taylor expansion
\begin{align*}
\alpha(t_1+\varepsilon)=&\alpha(t_1)+\dot\alpha(t_1)\varepsilon+\frac{1}{2}\ddot\alpha(t_1)\varepsilon^2
+\frac{1}{6}\dddot\alpha(t_1)\varepsilon^3+\cdots\\
=&\alpha(t_1)+\frac{1}{2-r}\dot\alpha(t_1)xL^{-1}+
\Big(-\frac{r}{2-r}\Psi_1\dot\alpha(t_1)+\frac{1}{2(2-r)^2}
\ddot\alpha(t_1)x^2\Big)L^{-2}+\cdots.
\end{align*}
Let us study the two cases

\begin{enumerate}
\item Now we set
\[
F=\alpha(t_1+\varepsilon)L^{1-r}=\alpha(t_1)
L^{1-r}+\frac{1}{2-r}\dot\alpha(t_1)x L^{-r}+\cdots.
\]
so that
\[
\bar X_s=-(1-r)\alpha(t_1)
\]
and
\[
\bar X(s)=\bar X-(1-r)s\alpha(t_1).
\]
When $r=1$ we get
\[
\xi(s)=\xi+s\alpha(t_1)
\]

\item In this case we have
\[
F=\alpha(t_1+\varepsilon)L^{2-r}=\alpha(t_1)
L^{2-r}+\frac{1}{2-r}\dot\alpha(t_1)x L^{1-r}+\cdots.
\]
which implies the following PDE for $\bar X$
\[
\bar X_s=\alpha(t_1)\bar
X_{t_1}-\frac{1-r}{2-r}\dot\alpha(t_1)\bar X
\]
whose solution is
\[
\bar X(s)=\alpha(t_1)^{\frac{1-r}{2-r}}f\Big(s+\int^{t_1}\frac{\d
t}{\alpha(t)}\Big),\quad\text{ with $f$ an arbitrary function }
\]
which leads to the symmetry
\[
\tilde{\bar X}=\dot T^{-\frac{1-r}{2-r}}\bar X(x,T(t_1),\bar t_1).
\]
When $r=1$ we get the following PDE for $\xi$
\[
\xi_s=\alpha\xi_{t_1}+\dot\alpha(t_1)x
\]
with general solution
\[
\xi(s)=-x\log\alpha+f\Big(s+\int^{t_1}\frac{\d t}{\alpha(t)}\Big)
\]
leading to the symmetry
\[
\tilde\xi=x\log\dot T+\xi(x,T(t_1),\bar t_1).
\]
\end{enumerate}

There are additional symmetries associated with the $\bar t_1$
flow. Now $\bar N=1$ and $n=-r$ and $n=1-r$  and there are two
different generators:
\[
\bar \alpha\big(\frac{\bar M}{-r\bar L^{-1}}\big)\bar
L^{1-r}\text{ and }\bar \alpha\big(\frac{\bar M}{-r\bar
L^{-1}}\big)\bar L^{-r}.
 \]
Notice  that
\[
\frac{\bar M}{-r\bar L^{-1}}=\bar t_1-\bar\varepsilon
\]
with
\[
\bar \varepsilon:=\frac{1}{r}X\bar L+\frac{2-r}{r}\bar
\Psi_1(X)\bar L^2-\cdots,
\]
we also have the following Taylor expansion
\begin{align*}
\bar \alpha(\bar t_1+\bar\varepsilon)=&\bar \alpha(\bar
t_1)-\dot{\bar\alpha}(t_1)\bar\varepsilon+
\frac{1}{2}\ddot{\bar\alpha}(t_1)\bar\varepsilon^2
+\cdots\\
=&\bar \alpha(\bar t_1)-\frac{1}{r}\dot{\bar \alpha}(\bar
t_1)X\bar L+ \Big(-\frac{2-r}{r}\bar
\Psi_1(X)\dot{\bar\alpha}(\bar t_1)+\frac{1}{2r^2}
\ddot{\bar\alpha}(\bar t_1)X^2\Big)\bar L^2+\cdots.
\end{align*}
The two generators are
\begin{align}
\bar F=&\bar \alpha(\bar
t_1)\bar L^{1-r},\label{gen1}\\
\bar F=&\bar \alpha(\bar t_1)\bar L^{-r}-\frac{1}{r}\dot{\bar
\alpha}(\bar t_1)X\bar L^{1-r}.\label{gen2}
\end{align}

To deal with the symmetries we recall that
\[\begin{aligned}
\frac{\d\bar\psi_{1-r}}{\d
s}\cdot\bar\psi_{1-r}^{-1}&=P_{1-r}\Ad_{\bar\psi_{1-r}}\bar F(\bar
L,\bar M)=P_{1-r}\bar F(\bar \ell,\bar m),\\
\frac{\d\bar\psi_{1-r}}{\d \bar
t_n}\cdot\bar\psi_{1-r}^{-1}&=P_{1-r}\Ad_{\bar\psi_{1-r}}\bar
L^{1-r-n}=P_{1-r}\bar \ell^{1-r+n},
\end{aligned}
\]

With the first generator given in \eqref{gen1} we have
\[
\frac{1}{1-r}\frac{\bar X_s}{\bar X_x}=\bar\alpha(\bar t_1),
\]
so that $\bar X$ is subject to the following PDE
\[
\bar X_s=(1-r)\bar\alpha(\bar t_1)\bar X_x
\]
whose solution is
\[
\bar X(s)=f(x+s(1-r)\bar\alpha(\bar t_1),t_1,\bar t_1),
\]
with $f$ an arbitrary function. Then, the corresponding symmetry
transformation is given by
\[
\tilde{\bar X}=\bar X(x+(1-r)\bar\alpha(\bar t_1),t_1,\bar t_1).
\]

For the second generator in \eqref{gen2} we find the following PDE
\[
\bar X_s=\bar\alpha(\bar t_1)\bar X_{\bar
t_1}-\frac{1-r}{r}\dot{\bar\alpha}(\bar t_1)\bar X_x
\]
whose general solution is
\[
\bar X(s)=f\Big(x\alpha(\bar t_1)^{\frac{1-r}{r}},s+\int^{\bar
t_1}\frac{\d t}{\alpha(t)}\Big)
\]
with $f$ an arbitrary  function. Proceeding as before we find a
new symmetry:
\[
\tilde{\bar X}=\bar X(\dot{\bar T}(\bar
t_1)^{-\frac{1-r}{r}}x,t_1,\bar T(\bar t_1))
\]
with $\bar T$ an arbitrary function on $\bar t_1$.

For the $r=1$ case  we may proceed as above, however notice that
in the $r=1$ case we have the dToda equation,
$(\exp(\xi_x))_x+\xi_{t_1\bar t_1}=0$, the interchange of $t_1$
and $\bar t_1$ leaves the equation invariant. Thus, the symmetries
are as the one derived already for the $F(L,M)$ generators but
replacing $t_1$ by  $\bar t_1$. Namely:
\begin{align*}
\tilde\xi=&\xi(x,t_1,\bar t_1)+\bar\alpha(\bar t_1),\\
\tilde\xi=&x\log\dot{\bar T}(\bar t_1)+\xi(x,t_1,\bar T(\bar t_1))
\end{align*}

We collect these results regarding the $r$-dToda equation in the
following
\begin{pro}
Given a solution $\bar X$ of the $r$-dToda equation
\[
\Big((\bar
X_x)^{-\frac{2-r}{1-r}}\Big)_x-\frac{1}{(1-r)r}\Big(\frac{\bar
X_{\bar t_1}}{\bar X_x}\Big)_{ t_1}=0,
\]
and arbitrary functions $\alpha(t_1)$, $T(t_1)$, $\bar\alpha(\bar
t_1)$ and $\bar T(\bar t_1)$ the following  functions are new
solutions of the $r$-dToda equation:
\begin{align*}
\tilde{\bar X}=& \bar X(x,t_1,\bar t_1)-(1-r)\alpha(t_1),\\
\tilde{\bar X}=& \dot T(t_1)^{-\frac{1-r}{2-r}}\bar X(x,T(t_1),\bar t_1),\\
\tilde{\bar X}=& \bar X(x+(1-r)\bar\alpha(\bar t_1),t_1,\bar t_1),\\
\tilde{\bar X}=& \bar X(\dot{\bar T}(\bar
t_1)^{-\frac{1-r}{r}}x,t_1,\bar T(\bar t_1)).
\end{align*}

\end{pro}

Finally, observe that the symmetries derived for the $r=1$ case,
i. e., the Boyer--Finley equation $(\exp(\xi_x))_x+\xi_{t_1\bar
t_1}=0$, are:
\begin{align*}
\tilde\xi=&\xi(x,t_1,\bar t_1)+\alpha(t_1),\\
\tilde\xi=&x\log\dot{T}(t_1)+\xi(x,T(t_1),\bar t_1),\\
\tilde\xi=&\xi(x,t_1,\bar t_1)+\bar\alpha(\bar t_1),\\
\tilde\xi=&x\log\dot{\bar T}(\bar t_1)+\xi(x,t_1,\bar T(\bar
t_1)).
\end{align*}
These symmetries are the well known, since 1986 by P. Olver,
conformal symmetries of the dToda equation.

\section{ On $t_2$ invariance and solutions for the potential
$r$-dDym and $r$-dmKP equations}

Here we study the $t_2$ invariance on the equations potential
$r$-dDym equation \eqref{PDEbarX} and using the Miura type map
corresponding solutions for the  $t_2$ invariant solutions of the
$r$-dmKP equation \eqref{r-dmKP}, which we may call $r$
dipersionless modified Boussinesq equation.

\subsection{$t_2$-invariance for the $r$-dDym equation}
 We analyze here solutions of the \eqref{PDEbarX}
which do not depend on one of the variables  $t_2$.
%\begin{itemize}
%  \item $t_1$-invariance. Let us suppose that $\bar X$ do not
%  depend on $t_1$, then \eqref{PDEbarX} simplifies to
%  \[
%  \bar X_{xt_2}=0
%  \]
%and therefore its general solution is of the form
%\[
%\bar X=f(x)+g(t_2)
%\]
%being $f$ and $g$ arbitrary functions of one variable.
Thus, \eqref{PDEbarX} simplifies to
\[
(1-r)\bar X_{t_1t_1}\bar X_x=(2-r)\bar X_{xt_1}\bar X_{t_1}
\]
which can be written as
\[
(\log(\bar X_{t_1}^{1-r}))_{t_1}=(\log(\bar X_x^{2-r}))_{t_1}
\]
so that
\[
\Big(\log\Big(\frac{\bar X_{t_1}^{1-r}}{\bar
X_x^{2-r}}\Big)\Big)_{t_1}=0.
\]
This last equation is equivalent to
\[
\frac{1}{k'(x)}\bar X_x=\bar X_{t_1}^{\frac{1-r}{2-r}}
\]
where $k'$ is the derivative of $k(x)$, an arbitrary function of
$x$. Thus, if we introduce the variable $\tilde x=k(x)$ we have
\[
\bar X_{\tilde x}=\bar X_{t_1}^{\frac{1-r}{2-r}}.
\]
This is a first order nonlinear (for $1-r\neq 0$) PDE, and we will
solve it by the the method of the complete solution. First observe
that a complete integral is
\[
\bar X(\tilde x,t_1;a,b)=a^{\frac{1-r}{2-r}}\tilde x+at_1+b.
\]
Set $b=f(a)$ so that we get a uni-parametric family of solutions
depending on arbitrary function $f$
\[
\bar X(\tilde x,t_1;a)=a^{\frac{1-r}{2-r}}\tilde x+at_1+f(a).
\]
To find the envelope of this solution we request
\[
\bar X_a=\frac{1-r}{2-r}a^{-\frac{1}{2-r}}\tilde x+t_1+f'(a)=0,
\]
which determines locally a solution of the form
\[
a:=a(\tilde x,y).
\]
Then,  the corresponding envelope is given by
\[
\bar
X(x,t_1)=a(k(x),t_1)^{\frac{1-r}{2-r}}k(x)+a(k(x),t_1)t_1+f(a(k(x),t_1)
\]
which depends on two arbitrary functions $k$ and $f$ on one
variable, being therefore a general solution.

For example, if $f=0$ we get
\[
a(\tilde x,t_1)=\Big(-\frac{1-r}{2-r}\frac{\tilde
x}{t_1}\Big)^{2-r}
\]
and the solution is
\begin{gather}\label{solutiondym1}
\bar X=\frac{\kappa(x)^{2-r}}{t_1^{1-r}},\quad \quad
\kappa(x):=(-1)^{1-r}\frac{(1-r)^{1-r}}{(2-r)^{2-r}}k(x),
\end{gather}
Observe that $\kappa$ may be considered is an arbitrary function.

We may get more general solutions    of the potential $r$-dDym
equation \eqref{PDEbarX} which are not $t_2$ invariant by applying
the symmetries given in Proposition \ref{symmetries-prddym}. We
have the following new solutions
\begin{align*}
\tilde{\bar X}=& \frac{\kappa(x)^{2-r}}{t_1^{1-r}}-(1-r)\alpha(t_2),\\
\tilde{\bar
X}=&-\frac{(1-r)(2-r)}{2(3-r)}\dot\alpha(t_2)(\alpha(t_2)+2t_1)+
\frac{\kappa(x)^{2-r}}{(t_1+ \alpha(t_2))^{1-r}},\\
\tilde{\bar X}=& -\frac{(1-r)(2-r)^2}{2(3-r)^2}t_1^2 \frac{\ddot
T(t_2)}{\dot T(t_2)}+\frac{\kappa(x)^{2-r}}{(\dot
T(t_2)t_1)^{1-r}}.
\end{align*}
The second and third family are non trivial solutions depending on
on two arbitrary functions of one variable each.

The case $f(a)=-\frac{2-r}{r}a^{-\frac{r}{2-r}}$ leads to the
equation
\[
\alpha^2+\frac{1-r}{2-r}\tilde x\alpha+t_1=0,\quad
A:=a^{-\frac{1}{2-r}}
\]
whose solution is
\[
\alpha=-\frac{1-r}{2-r}\frac{\tilde
x}{2}\pm\sqrt{\frac{(1-r)^2}{(2-r)^2}\frac{\tilde x^2}{4} -t_1}
\]
and the solution is

\[
\bar X=\Big(\frac{\tilde
x}{\alpha}+\frac{t_1}{\alpha^2}-\frac{2-r}{r}\alpha\Big)\alpha^r.
\]

Finally, if we consider $f(a)=-\frac{2-r}{1+r}
a^{\frac{2-r}{1+r}}$ and introduce the functions
\[
F:=\sqrt[3]{108t_1+12\sqrt{3}\sqrt{4\frac{(1-r)^3}{(2-r)^3}\tilde
x^3+t_1^2}},\quad
\alpha:=-\frac{1}{6}F+2\frac{1-r}{2-r}\frac{\tilde x}{F}
\]
a solution is
\[
\bar X=\Big(\frac{\tilde
x}{\alpha}+\frac{t_1}{\alpha^2}-\frac{2-r}{1+r}\alpha\Big)\alpha^r.
\]
%\end{itemize}

\subsection{Solutions of the $r$-dmKP through Miura map} As an example we shall use the Miura map
\[
-\frac{1}{(1-r)(2-r)}\bar
X_{t_1}(X(x,t_1,t_2),t_1,t_2)=u(x,t_1,t_2)
\]
to get a solution of the $r$-dmKP equation from the solution of
the $r$-dDym equation as given in \eqref{solutiondym1}, which is
$t_2$ independent. The inverse function $X$ is given, in this case
in explicit form , by
\[
X(x,t_1,t_2)=k^{-1}\bigg(\Big(
\frac{xt_1^{1-r}}{C}\Big)^{\frac{1}{2-r}}\bigg),\quad
 C=(-1)^{1-r}\frac{(1-r)^{1-r}}{(2-r)^{2-r}},
\]
and as
\[
\bar X_{t_1}=-C(1-r)\frac{k(x)^{2-r}}{t_1^{2-r}}
\]
we get the corresponding solution of the $r$-dmKP equation
\eqref{r-dmKP}
\begin{gather}\label{sol_not2_dmkP}
u=\frac{1}{2-r}\frac{x}{t_1}.
\end{gather}
The corresponding solution of the potential $r$-dmKP \eqref{mKP}
is
\[
\Psi_1=-\frac{1}{2(2-r)}\frac{x^2}{t_1},
\]
and applying Proposition \ref{symmeries-rdmkp} we get the
following solutions
\begin{align*}
\tilde\Psi_1=&-\frac{2-r}{3-r}\dot\alpha(t_2)t_1-\frac{1}{2(2-r)}\frac{(x+(1-r)\alpha(t_2))^2}{t_1},\\
\tilde\Psi_1= &-\frac{1}{3-r}\dot\alpha(t_2) x-
\frac{(2-r)^2}{2(3-r)^2}\ddot\alpha(t_2)
t_1^2\\&-\frac{2-r}{6(3-r)^2}((1-r)\dot\alpha(t_2)^2+(2-r)\alpha(t_2)
\ddot\alpha(t_2))(3t_1+\alpha(t_2))\\
&-\frac{1}{2(2-r)}\frac{(x+\frac{(1-r)(2-r)}{2(3-r)}\dot\alpha(t_2)(\alpha(t_2)+2t_1))^2}{t_1
+\alpha(t_2)},\\[8pt]
\tilde \Psi_1=&-\frac{2-r}{2(3-r)}\frac{\ddot T(t_2)}{\dot
T(t_2)}xt_1-\frac{(2-r)^3}{2(3-r)^3}\Big(\frac{1}{3}\frac{\dddot
T(t_2)}{\dot T(t_2)}-\frac{1+r}{4}\,\frac{\ddot T(t_2)^2}{\dot
T(t_2)^2}\Big)t_1^3-\frac{1}{2(2-r)}\frac{x^2}{t_1}.
\end{align*}

Observing that
\[
\bar X_{t_1}=\bar X_a a_{t_1} +a
\]
and recalling that $\bar X_a=0$ we get the corresponding solution
\[
u=-\frac{1}{(1-r)(2-r)}a(X,t_1,t_2)
\]
of the potential $r$-dmKP equation.

\section{Twistor equations}

Previously we have introduced the Lax and Orlov functions as the
following canonical transformations of the pair $p,x$:
\begin{equation*}
\begin{aligned}
L&=\Ad_{\psi_<\cdot\exp t}p,&\bar \ell&=\Ad_{\psi_>\cdot\exp \bar
t}p,&\bar L&=\Ad_{\psi_\geqslant\cdot\exp \bar t}p,\\
M&:=\Ad_{\psi_<\cdot\exp t}x,&\bar m&:=\Ad_{\psi_>\cdot\exp \bar
t}x,&\bar M&:=\Ad_{\psi_\geqslant\cdot\exp \bar t}x.
\end{aligned}
\end{equation*}
Thus, they satisfy
\[
\{L,M\}=L^r,\quad \{\bar \ell,\bar m\}=\bar\ell^r,\quad \{\bar
L,\bar M\}=\bar L^r.
\]
Another important functions  are
\[
P:=\Ad_{h}p,\quad Q:=\Ad_h x,\quad \bar P:=\Ad_{\bar h}p,\quad
\bar Q:=\Ad_{\bar h} x
\]
for which  we have
\[
\{P,Q\}=P^r,\quad \{\bar P,\bar Q\}=\bar P^r.
\]
These functions result from the canonical transformation of the
$p,x$ variables generated by the initial conditions $h,\bar h$ of
the factorization problem \eqref{factorization}.

We are ready for

\begin{pro}
For any solution $\psi_<$ and $\psi_\geqslant$ of the
factorization problem \eqref{factorization} the following twistor
equations hold
\begin{equation}\label{twistor}
\begin{split}
P(L,M)&=\bar P(\bar L,\bar M),\\
Q(L,M)&=\bar Q(\bar L,\bar M).
\end{split}
\end{equation}
\end{pro}
\begin{proof}
The factorization problem solve
\[
\psi_<\cdot\exp(t)\cdot h=\psi_\geqslant\cdot\exp(\bar t)\cdot\bar
h
\]
implies for any function $F(p,x)$
\[
\Ad_{\psi_<\cdot\exp(t)}\Ad_{h}F(p,x)=\Ad_{\psi_\geqslant\cdot\exp(\bar
t)}\Ad_{\bar h}F(p,x).
\]
Hence,
\[
\Ad_{\psi_<\cdot\exp(t)}F(P(p,x),Q(p,x))=\Ad_{\psi_\geqslant\cdot\exp(\bar
t)}F(\bar P(p,x),\bar Q(p,x))
\]
and therefore
\[ F(P(L,M),Q(L,M))=F(\bar P(\bar L,\bar M),\bar Q(\bar L,
\bar M))
\]
as desired.
\end{proof}

%If we take $\bar h =\text{id}$ and keep $\bar t_n=0$, i. e., we
%are dealing with the $r$-dmKP type for $\psi_<$ and with the
%$r$-dDym hierarchy when $\psi_{1-r}$ is considered as well. Then,
%the twistor conditions can be written as
%\[
%\begin{aligned}
%P(L,M)&=\bar L|_{\bar t=0},\\
%Q(L,M))&= \bar M|_{\bar t=0}
%\end{aligned}
%\]
%which in particular imply
%\begin{equation}\label{twistor-simple}
%\begin{aligned}
%P(L,M)_{<1}&=0,\\
%Q(L,M))_{<0}&=0.
%\end{aligned}
%\end{equation}
%We need to explain that given a Laurent series $f(p,x)$ the
%Laurent series $f_{<n}(p,x)$ is the truncated series with powers
%in $p$ of order less than $n$.

% The $S$- function is as in
%\eqref{Stfix} and if we assume that $f+S\in\g_\geqslant$ we have
% \begin{multline}
%j=f(L)_\geqslant+t_2\Big(p^{3-r}+(3-r)up^{2-r}+(3-r)\Big(v+\frac{2-r}{2}u^2\Big)p^{1-r}\Big)\\+t_1
%(p^{2-r}+(2-r)up^{1-r})+\frac{x}{1-r}p^{1-r}.
%\end{multline}
%As
%\[
%P_p=(3-r)p^{-r}\Big(p^2+(2-r)up+(1-r)\Big(v+\frac{2-r}{2}u^2\Big)\Big)
%\]
%and its zeroes are
%\[
%p_\pm=-\frac{2-r}{2}u\pm\Delta,\quad
%\Delta:=\sqrt{\frac{(2-r)r}{4}u^2-(1-r)v}
%\]
%
%Now,
%\[
%j_p\big|_{p=p_\pm}= (f(L)_\geqslant)_p\big|_{p=p_\pm}+ (t_1
%(2-r)(p_\pm+(1-r)u)+x)p_\pm^{-r}=0
%\]
%can be written as
%\[
%p_\pm^{r}(f(L)_\geqslant)_p\big|_{p=p_\pm}-
%\frac{(2-r)r}{2}ut_1+x\pm t_1 (2-r)\Delta=0
%\]

\section*{Acknowledgements}

The author is in debt with Luis Mart{\'\i}nez Alonso and Elena Medina
for several discussions. Partial economical support from
Direcci\'{o}n General de Ense\~{n}anza Superior e
Investigaci\'{o}n Cient\'{\i}fica n$^{\mbox{\scriptsize
\underline{o}}}$ BFM2002-01607 is also acknowledge.

\end{document}